\documentclass[twocolumn]{aastex63}
\usepackage{graphicx}
\usepackage{amssymb}
\usepackage{amsmath}
\usepackage{epstopdf}
\usepackage{calc}
\shorttitle{1.4\,GHz Source Counts} \shortauthors{Matthews et al.}

\begin{document}

\pdfsuppresswarningpagegroup=1
\maxdeadcycles=1000

\title{Source Counts Spanning Eight Decades of Flux Density at 1.4\,GHz}

\correspondingauthor{Allison M. Matthews} \email{amm4ws@virginia.edu}

\author[0000-0002-6479-6242]{A.~M.~Matthews}
  \affiliation{Department of Astronomy, University of
  Virginia, Charlottesville, VA 22904, USA} \affiliation{National Radio
  Astronomy Observatory, 520 Edgemont Road, Charlottesville, VA 22903, USA}

\author[0000-0003-4724-1939]{J.~J.~Condon}
  \affiliation{National Radio Astronomy Observatory, 520
  Edgemont Road, Charlottesville, VA 22903, USA}

\author[0000-0001-7363-6489]{W.~D.~Cotton}
  \affiliation{National Radio Astronomy Observatory, 520
  Edgemont Road, Charlottesville, VA 22903, USA}

\author[0000-0003-2716-9589]{T.~Mauch}
  \affiliation{South African Radio Astronomy Observatory
  (SARAO), 2 Fir Street, Black River Park, Observatory, 7925, South Africa}

\begin{abstract}
  Brightness-weighted differential source counts $S^2 n(S)$ spanning
  the eight decades of flux density between $0.25\,\mu\mathrm{Jy}$ and
  25\,Jy at 1.4\,GHz were measured from (1) the confusion brightness
  distribution in the MeerKAT DEEP2 image below $10\,\mu\mathrm{Jy}$,
  (2) counts of DEEP2 sources between $10\,\mu\mathrm{Jy}$ and
  $2.5\,\mathrm{mJy}$, and (3)
  counts of NVSS sources stronger than $2.5\,\mathrm{mJy}$.  We
  present our DEEP2 catalog of $1.7 \times 10^4$ discrete sources
  complete above $S = 10\,\mu\mathrm{Jy}$ over $\Omega = 1.04
  \,\mathrm{deg}^2$.
  The brightness-weighted counts converge as $S^2 n(S) \propto
  S^{1/2}$ below $S = 10\,\mu\mathrm{Jy}$, so $>99$\% of the $\Delta
  T_\mathrm{b} \sim 0.06\,\mathrm{K}$ sky brightness produced by
  active galactic nuclei and $\approx96$\% of the $\Delta T_\mathrm{b}
  \sim 0.04\,\mathrm{K}$ added by star-forming galaxies has been
  resolved into sources with $S \geq 0.25\,\mu\mathrm{Jy}$.
  The $\Delta T_\mathrm{b} \approx 0.4\,\mathrm{K}$ excess brightness
  measured by ARCADE 2 cannot be produced by faint sources smaller
  than $\approx 50\,\mathrm{kpc}$ if they cluster like galaxies.
\end{abstract}

\keywords{galaxies: evolution -- galaxies: star formation -- galaxies:
  statistics -- radio continuum: galaxies}



\section{Introduction}

There have been persistent discrepancies in the faintest direct
source counts at $S_{\mathrm{1.4\,GHz}} < 100\,\mu$Jy \citep[see][for a
review and compilation of previous source counts]{dezotti10}, far exceeding
the errors caused by Poisson fluctuations and clustering uncertainties
\citep{owen08, heywood13}.  Direct counts of faint radio sources rely
primarily on high angular-resolution images, and must account for possible
``missing'' resolved sources whose peak flux density falls below the surface
brightness sensitivity of the image.  Corrections on the source counts due
to these missing sources depend on the highly uncertain, intrinsic angular
source size distribution of faint radio sources \cite{bondi08}. The large
uncertainty in these resolution corrections propagates into the integrated
flux measurements, source counts, and number of missing sources due to
limited surface brightness sensitivty.  This effect is further magnified by
the steep slope of differential source counts $n(S) \propto S^{-5/2}$, which
exacerbates flux density overestimates and leads to higher counts at faint
flux densities.

Following the pioneering $P(D)$ method of \cite{scheuer57}, radio
astronomers have utilized confusion to measure accurate source counts
\citep[see][for recent examples]{condon12, vernstrom14}. A
low resolution image ensures that all faint radio galaxies appear as point
sources---eliminating the need for uncertain resolution
corrections. The term ``confusion'' means fluctuations
in sky brightness caused by multiple faint sources inside the point-source
response. Historically, confusion was described by the probability
distribution $P(D)$ of pen deflections of magnitude $D$ on a chart-recorder
plot of detected power \citep{scheuer57}.  The analog of the deflection $D$
in a modern image is the sky brightness expressed as a peak flux density
$S_\mathrm{p}$ in units of flux density per beam solid angle, so a
$P(D)$ distribution is the same as a $P(S_\mathrm{p})$ distribution.

Low resolution, confusion-limited images offer an independent
  way of measuring faint radio source counts and are free from uncertain
  angular size corrections.  While unable to determine properties of
  individual
  galaxies, confusion studies are able to constrain source counts of
  the radio population far below the noise and do not require
  multi-wavelength
  cross-identifications as priors \citep[unlike source counts
    measured from ``stacking'', e.g.][]{mitchell14}. 

The differential source count $n(S) dS$ at frequency $\nu$ is the
number of sources per steradian with flux densities between $S$ and $S
+ dS$.  The Rayleigh-Jeans sky brightness temperature $d T_\mathrm{b}$
per decade of flux density added by these sources is
\begin{equation}\label{eq:tb}
\Biggl[ \frac {d \,T_\mathrm{b}} {d \log(S)}\Biggr] =
\Biggl[ \frac {\ln(10)\,c^2} { 2 k_\mathrm{B} \nu^2} \Biggr]  S^2 n(S)~,
\end{equation}
where $k_\mathrm{B} \approx 1.38 \times 10^{-23}\mathrm{\,J\,K}^{-1}$.
Can one or more ``new'' populations of radio sources fainter than
$0.25\,\mu\mathrm{Jy}$ make comparable contributions to the sky
brightness at 1.4\,GHz?  The ARCADE 2 instrument measured the absolute
sky temperature at frequencies from $\nu = 3-90$ GHz, and
\citet{fixsen11} reported finding an excess power-law brightness
temperature
\begin{equation}
  \biggl(\frac {T_\mathrm{b}} {\mathrm{K}} \biggr)
  = (24.1 \pm 2.1)\left(\frac{\nu}{\nu_0}\right)^{-2.599\pm0.036}
\end{equation}
from 22\,MHz to 10\,GHz, where $\nu_0 = 310$ MHz. Removing the
contribution from known populations of extragalactic sources leaves
\begin{equation}
  \biggl( \frac {\Delta T_\mathrm{b}} {\mathrm{K}} \biggr)
    = (18.4 \pm 2.1)\left(\frac{\nu}{\nu_0}\right)^{-2.57\pm0.05},
\end{equation}
\citep{seiffert11}.  Possible explanations for this large excess fall into
three categories: 1) the excess was overestimated owing to the limited sky
coverage of ARCADE 2 and the zero-point levels of low frequency radio
maps may be inaccurate, 2) the excess is primarily smooth
emission from our Galaxy, 3) the excess is primarily extragalactic,
making it the only photon background that does not agree with
published source counts dominated by radio galaxies and star-forming
galaxies.  \citet{vernstrom11} and \citet{seiffert11} explored the
possibility of a new source population contributing an extra ``bump''
to the source counts at flux densities $<10\,\mu{\rm Jy}$.  In order
to match the ARCADE 2 excess background, this hypothetical new
population must add $\Delta T_\mathrm{b} \sim 0.4\,\mathrm{K}$ to the
sky brightness at 1.4\,GHz.  \citet{condon12} showed that the
brightness-weighted counts $S^2 n(S)$ of this new population must peak
at flux densities below $S_{1.4\,\mathrm{GHz}} = 0.1\,\mu{\rm Jy}$ to
be consistent with their observed $P(D)$ distribution.

This paper presents 1.4\,GHz brightness-weighted source counts $S^2
n(S)$ covering the eight decades of flux density between $S = 0.25
\,\mu\mathrm{Jy}$ and $S = 25\,\mathrm{Jy}$ based on the very
sensitive $\nu = 1.266\,\mathrm{GHz}$ MeerKAT DEEP2 sky image
\citep{mauch20} confusion brightness distribution between $S =
0.25\,\mu\mathrm{Jy}$ and $S = 10\,\mu\mathrm{Jy}$, the DEEP2
discrete-source catalog from $S = 10\,\mu\mathrm{Jy}$ to $S =
2.5\,\mathrm{mJy}$, and on the 1.4~GHz NRAO VLA Sky Survey
\citep[NVSS]{condon98} catalog above $S = 2.5\,\mathrm{mJy}$.
Nearly all of these sources are extragalactic. We present the
  first complete catalog of disrete sources with $S > 10\,\mu$Jy in
  the DEEP2 field. While \cite{mauch20} derived the best-fitting
power-law source counts to describe the DEEP2 $P(D)$ distribution, the
actual source counts do not follow a simple power law.  To improve
upon the \citet{mauch20} fit, we allowed the source counts to be any
continuous function.  We further explore the possibility of ``new''
populations of faint extragalactic sources contributing to the total
radio background, and constrain the lower limit to the number of such
sources adding $\Delta T_\mathrm{b} \sim 0.4\,\mathrm{K}$ to the
sky brightness at 1.4\,GHz remaining consistent with the DEEP2 $P(D)$
distribution.

The data used to construct the source counts across eight decades
is presented in Section \ref{sec:data}.  The radio sky simulations
needed to constrain the source counts from the $P(D)$ distribution
and derive confusion corrections to the discrete counts is detailed
in Section \ref{sec:model}.  Statistical source counts with
$0.25\mu\mathrm{Jy} < S < 10\,\mu\mathrm{Jy}$ estimated from the $P(D)$
confusion distribution are reported in Section \ref{sec:pofd}. Section
\ref{sec:deep2cat} presents the complete $S > 10\,\mu\mathrm{Jy}$
discrete source catalog from the DEEP2 field, and Section \ref{sec:deep2count}
presents the source counts derived from this catalog. Differential
source counts for the NVSS catalog are calculated in Section
\ref{sec:nvsscount}. The contributions of SFGs and AGNs to the 1.4\,GHz
sky background and constraints on ``new'' populations of faint sources
explaining the ARCADE 2 radio excess are described in Section
\ref{sec:discussion}. Section \ref{sec:summary} summarizes this work.

Absolute quantities in this paper were calculated for a $\Lambda$CDM
universe with $H_0 = 70\,\mathrm{km\,s}^{-1} \mathrm{\,Mpc}^{-1}$ and
$\Omega_\mathrm{m} = 0.3$ using equations in \citet{condon18}. Our
spectral-index sign convention is $\alpha \equiv + d\, \ln S / d\,\ln
\nu$.

\section{Data}\label{sec:data}

\subsection{The MeerKAT DEEP2 Field}

The 1.266\,GHz DEEP2 image
\citep{mauch20}
covers the $\Theta_{1/2} = 69\farcm 2$
diameter half-power circle of the MeerKAT primary beam centered on
J2000 $\alpha = 04^\mathrm{h}\,13^\mathrm{m}\, 26\,\fs 4$, $\delta =
-80^\circ\, 00'\,00''$. Its point-source response is a $\theta_{1/2} =
7\,\farcs 6$ FWHM Gaussian, and the rms noise is $\sigma_\mathrm{n} =
0.56 \pm 0.01 \,\mu\mathrm{Jy\,beam}^{-1}$ at the pointing center
(Table~\ref{tab:obs}).
The wideband DEEP2 image is the average of 14 narrow subband
images weighted to maximize the signal-to-noise ratio (SNR) of sources
with spectral index $\alpha = -0.7$ (Table \ref{tab:subbands}).
The dirty DEEP2 image was CLEANed down to residual peak flux density
$S_\mathrm{p} = 5\,\mu\mathrm{Jy\,beam}^{-1}$.

%
%
%
\begin{deluxetable}{l l}
  \caption{DEEP2 survey parameters}
  \label{tab:obs}
  \tablehead{
    Parameter & Value 
    }
\startdata
Right Ascension (J2000) & \phantom{$-$}04:13:26.4   \\
Declination (J2000)     & $-$80:00:00\phantom{.4}    \\
Primary FWHM $\Theta_{1/2}$ & $69\farcm2$ \\
Solid angle $\Omega_{1/2}$                & $1.04$\,deg$^2$  \\
Synthesized FWHM $\theta_{1/2}$           & $7\,\farcs6$ \\
Central rms noise $\sigma_{\mathrm{n}}$ & $0.56\pm 0.01 \,\mu\mathrm{Jy\,beam}^{-1}$  \\
$P(D)$ on-sky noise $\sigma_{\mathrm{n}}$ & $0.57\pm 0.01\,\mu\mathrm{Jy\,beam}^{-1}$  \\
\enddata
\end{deluxetable}

The DEEP2 image is strongly confusion limited, so we could not treat
its position and flux-density error distributions
analytically. Therefore we created radio sky simulations
(Section~\ref{sec:model}) to model the statistical source counts
consistent with the confusion brightness distribution, refine our
catalog of discrete DEEP2 sources, and correct our counts of the
faintest sources
(Section~\ref{sec:deep2cat}).

\subsection{NRAO VLA Sky Survey}

The 1.4~GHz NRAO VLA Sky Survey (NVSS) \citep{condon98} imaged the
entire sky north of J2000 $\delta = -40\degr$ with $\theta_{1/2} =
45\arcsec$ FWHM resolution and ${\sigma_\mathrm{n} \approx 0.45
\mathrm{~mJy~beam}^{-1}}$ rms noise.  The NVSS catalog lists source
components as Gaussian fits to significant peaks in the NVSS images.
From it we selected the 1117067 components with $S \geq 2.5$~mJy in
the $\Omega \approx 7.016$~sr solid angle with absolute
Galactic latitude $\vert b \vert
\geq 20^\circ$.
In Section \ref{sec:nvsscount} we detail how we derived the
NVSS direct source counts above 2.5\,mJy.

\section{The Radio Sky Simulations}\label{sec:model}

We produced computer simulations of the 1.266\,GHz DEEP2 image (along with
mock catalogs) to calculate source counts below $10\,\mu\mathrm{Jy}$,
assess the quality of the algorithms (e.g. our source finding algorithm)
used on the real data, and to derive
corrections and uncertainties for the discrete source catalog between
$10\,\mu\mathrm{Jy} < S < 2.5\,\mathrm{mJy}$.

The confusion brightness distribution can be calculated analytically only
for scale free power-law differential source counts of the form $n(S)
\propto S^{-\gamma}$ \citep{condon74}.  Likewise, population-law
biases in counts of
faint discrete sources can easily be estimated only in the power-law
count approximation \citep{murdoch73}.
The actual source counts 
 are not well approximated by a single
 power law near $S \sim 10\,\mu \mathrm{Jy}$
 because that flux density corresponds to the bend in the
SFG luminosity function of sources at $z \sim 1$
\citep[fig.~11]{condon12}, so we used computer simulations of the 1.266\,GHz DEEP2
image to estimate statistical source counts below $10\,\mu\mathrm{Jy}$
from the DEEP2 image brightness distribution and to correct for biases
in the
DEEP2 discrete-source counts above $10\,\mu\mathrm{Jy}$.  We simulated
only point sources because the measured median angular diameter
$\langle \phi \rangle \approx 0\,\farcs3$ of real $\mu$Jy sources
\citep{cotton18} is much smaller than the DEEP2 restoring beam
FWHM and only $\sim 0.2$\% of the DEEP2 sources stronger than $S =
10\,\mu\mathrm{Jy}$ are clearly resolved
(Section~\ref{sec:extendedsources}).  The simulated sources all have
spectral index $\alpha = -0.7$, the median spectral index of
extragalactic sources \citep{condon84}.
Varying $\alpha$ by $\pm 0.14$, the rms width of the observed
spectral-index distribution of faint sources, changes the 1.266\,GHz
flux densities of the simulated sources by only $\mp 1$\%.

The input for each simulation is an arbitrary user-specified 1.266\,GHz
source count $n(S)$.  In every flux-density bin of width $\Delta
\log(S) = 0.001$ the actual number of simulated sources is chosen by a
random-number generator sampling the Poisson distribution whose mean
matches the input $n(S)$.  The sources are scattered randomly throughout
the DEEP2 half-power circle.
The real $\mu\mathrm{Jy}$ sources in DEEP2 are nearly all
extragalactic and very distant (median redshift $\langle z \rangle
\sim 1$), so they are spread out over a radial distance range $\Delta
z \sim 1$ much larger than the galaxy correlation length and their sky
distribution is quite random and isotropic \citep{benn95,condon18},
unlike the visibly clustered sky distribution of nearby optically
selected galaxies.  In addition, clustering appears to have little
effect on FIR, millimeter, and radio confusion brightness distributions
observed
with resolutions close to
the DEEP2 restoring beam diameter \citep{bet17}.

\begin{deluxetable}{r c c c}
  \caption{DEEP2 imaging subband frequencies and weights}
  \label{tab:subbands}
  \tablehead{
    Subban{\rlap d} & $\nu_i$ & $\sigma_i$ & $w_i$ for \\
    numbe{\rlap r} & (MHz) & ($\mu\mathrm{Jy~beam}^{-1}$) & 
    max SNR 
    }
\startdata
 $i=1$ & \hphantom{1}908.040 &  4.224 &   0.0378 \\
     2 & \hphantom{1}952.340 &  5.044 &   0.0248 \\
     3 & \hphantom{1}996.650 &  3.196 &   0.0580 \\
     4 &   1043.460 &  2.882 &    0.0669 \\
     5 &   1092.780 &  2.761 &    0.0683 \\
     6 &   1144.610 &  2.580 &    0.0733 \\
     7 &   1198.940 &  4.203 &    0.0259 \\
     8 &   1255.790 &  3.981 &    0.0271 \\
     9 &   1317.230 &  1.851 &    0.1170 \\
    10 &   1381.180 &  1.643 &    0.1389 \\
    11 &   1448.050 &  1.549 &    0.1463 \\
    12 &   1519.940 &  1.871 &    0.0938 \\
    13 &   1593.920 &  2.888 &    0.0368 \\
    14 &   1656.200 &  1.850 &    0.0850 \\
\enddata
\tablecomments{Column 1 is the subband number $i$, column 2 the
  subband central frequency $\nu_i$, column 3 the rms noise $\sigma_i$
  in the subband image, and column 4 is the subband image weight $w_i$
  used to produce the wideband DEEP2 image with the highest
  signal-to-noise ratio (SNR) for sources with spectral index
  $\alpha = -0.7$.}
\end{deluxetable}

\begin{figure}[!htb]
  \centering
  \includegraphics[width=0.45\textwidth,trim = {1.cm 5.5cm 7cm 13cm},clip]
  {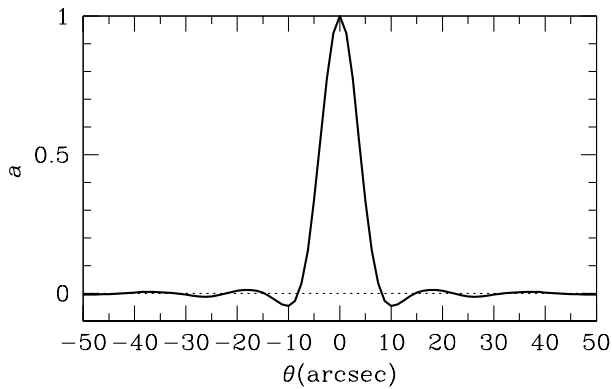}
  \caption{The SNR-weighted DEEP2 dirty beam, cut along the maximum
    sidelobes direction, has a 5\% negative
    sidelobe and a 1\% positive sidelobe.  Abscissa: Offset from the
    beam center (arcsec) Ordinate: Dirty beam power profile
    $a$.
  \label{fig:dirtybeam}}
\end{figure}

The simulations also reproduce the DEEP2 observational effects and
imaging processes described by \citet{mauch20}.  The simulated image
replicates CLEANing by representing each source as the sum of two
components: (1) a component whose brightness distribution is the DEEP2
dirty beam and whose peak flux density is the lesser of the input
source flux density or subband CLEAN threshold plus (2) a
CLEAN component whose brightness distribution matches the circular Gaussian
restoring beam
and whose amplitude
is the difference between the input source flux density and
the subband CLEAN threshold. The subband CLEAN
  threshold was determined from the wideband value of $5\,\mu$Jy and
  scaled to the subband central frequency assuming a spectral index of
  $\alpha = -0.7$.  The dirty beam used for each subband of the
simulation is the actual DEEP2 subband dirty beam, which is
  nearly circular and does not have strong diffraction spikes since
  the MeerKAT antennae
are not distributed along straight arms.  The first negative
sidelobe of the weighted dirty beam is at the $\sim$5\% level
(Figure~\ref{fig:dirtybeam}).  A 5\,$\mu$Jy residual leaves a $\sim
0.25\,\mu$Jy negative ring in the image, which is less than half the
rms sky noise in the $P(D)$ region. The first positive sidelobe of the
dirty beam is at the $\sim$1\% level.

\begin{figure*}[!ht]
  \centering
  \gridline{\hspace{-1cm}\fig{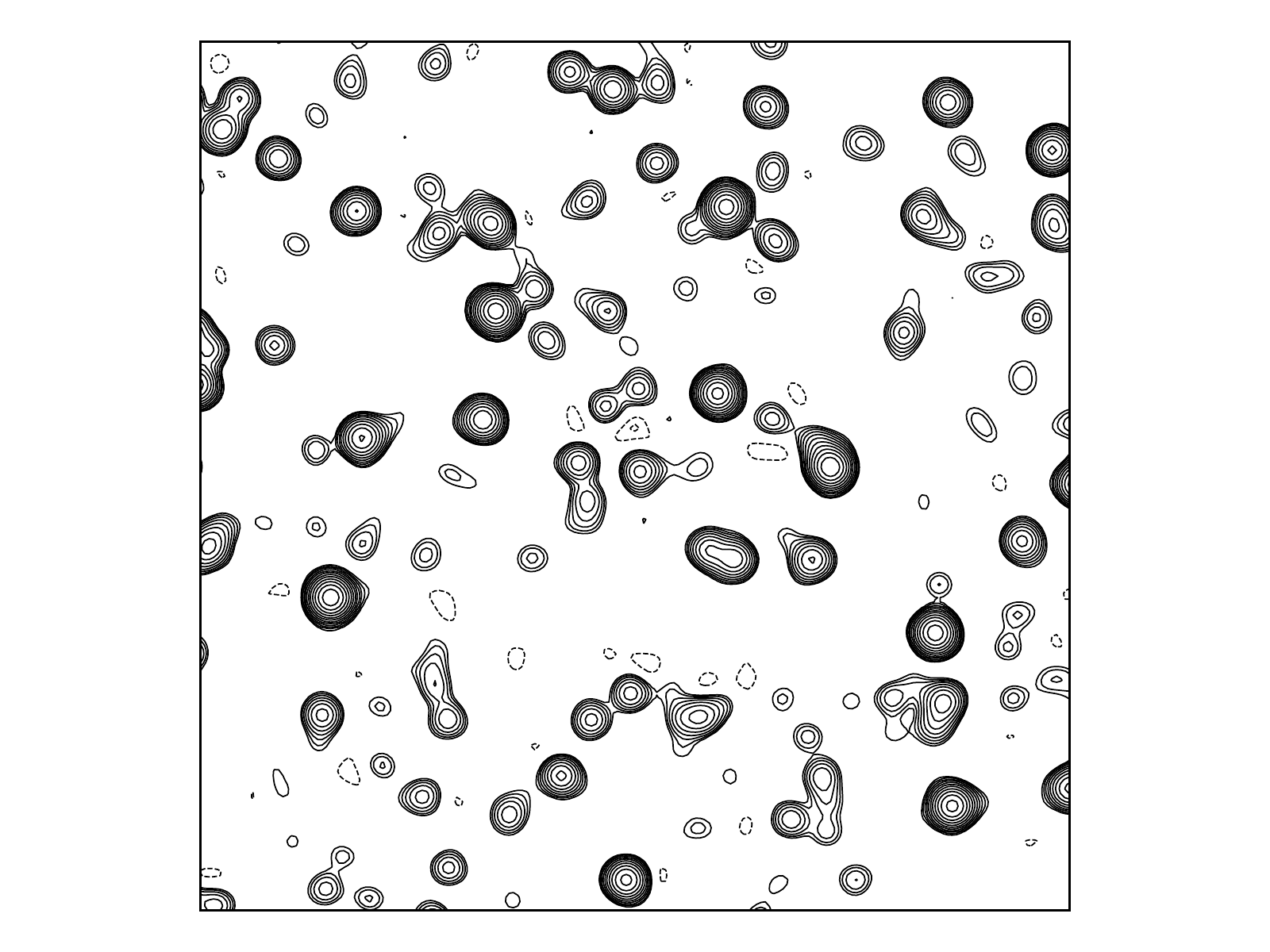}{0.6\textwidth}{}\hspace{-1.5cm}
    \fig{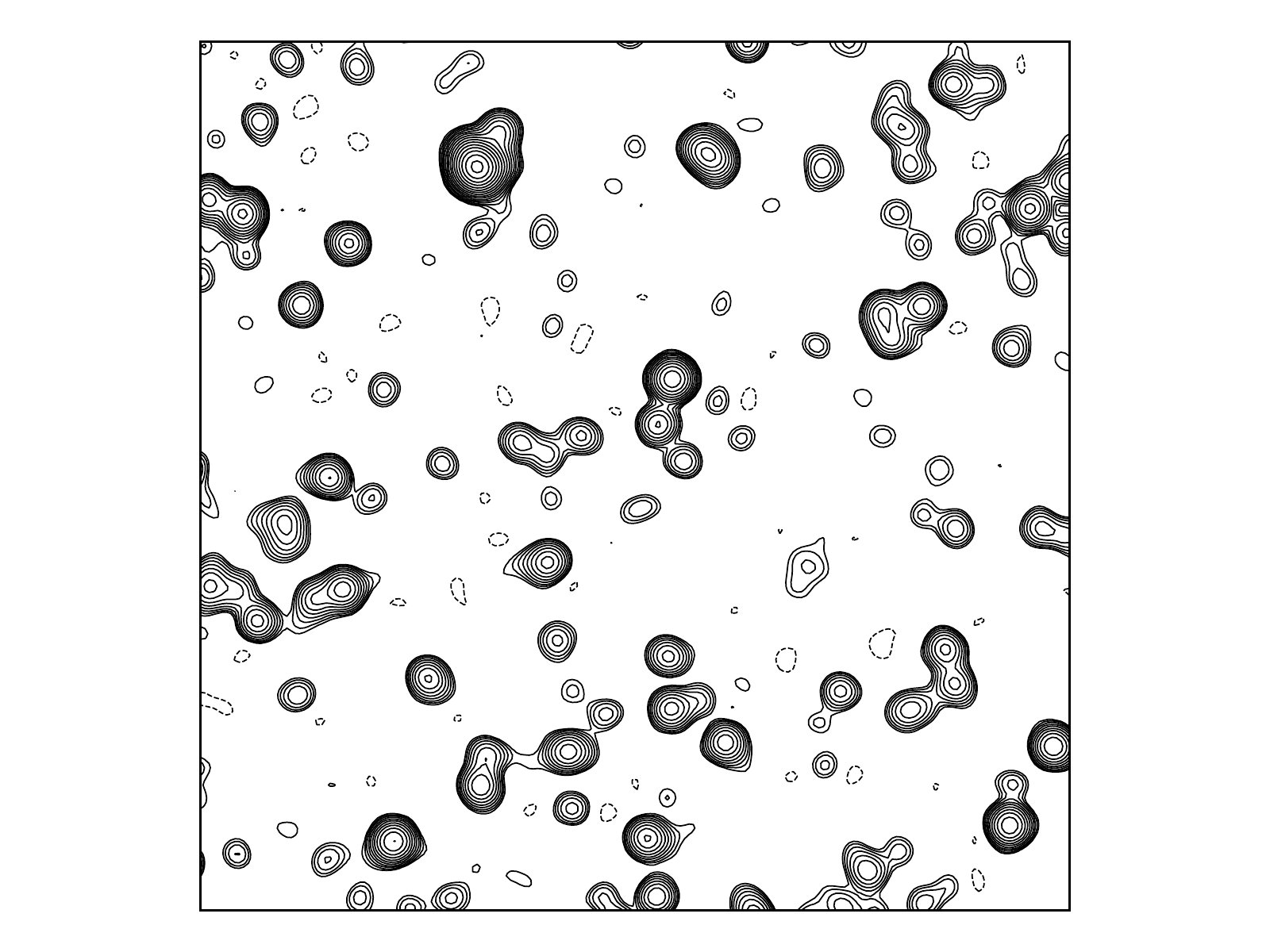}{0.6\textwidth}{}\hspace{-1cm}}
  \caption{These contour plots compare $4.2\arcmin \times 4.2\arcmin$
    regions of one DEEP2 simulation (left panel) and the actual DEEP2
    image (right panel).  Contours are drawn at 1.266\,GHz
    brightness levels
    $S_\mathrm{p} = \pm 
   (2^{1.5}, 2^2, 2^{2.5},
    ...) \,  \mu\mathrm{Jy\,beam}^{-1}$.}
  \label{fig:contours}
\end{figure*}

The simulation generates sources with spectral index $\alpha = -0.7$
and combines the subband images with the weights listed in column 4 of
Table \ref{tab:subbands}. To incorporate the DEEP2 primary beam
attenuation, the simulated subband images were multiplied by the
frequency-dependent MeerKAT primary beam specified by equations 3 and
4 in \citet{mauch20}. After multiplying by the primary beam
attenuation, the simulation adds to each pixel of the wideband image a
randomly generated sample of Gaussian noise.  The noise in an
aperture-synthesis image has the same $(u,v)$-plane coverage as the
signal, so the DEEP2 image noise is smoothed by the same dirty beam
(Figure~\ref{fig:dirtybeam}).  To duplicate this smoothing, the
simulation convolved the pixel noise distribution with the dirty beam.
The rms amplitude of the convolved noise  was set to match the
observed rms noise
in the actual DEEP2 image prior to
correction for primary beam attenuation.

Finally, this simulated image must be divided by primary beam
attenuation to yield a simulated sky image.  Figure~\ref{fig:contours}
compares $4\farcm2 \times 4\farcm2$ ($200\times200$ square pixels,
each $1\farcs25$ on a side) cutouts from one simulated sky image with
the actual DEEP2 sky image to show that the simulated image looks like
the real image.

\section{The DEEP2 $P(D)$ Distribution}\label{sec:pofd}

\subsection{Observed $P(D)$ Distribution}\label{sec:obspofd}
\begin{figure*}
  \centering
  \includegraphics[width=0.85\textwidth,trim = {0cm 0.5cm 1cm 8cm},clip]
    {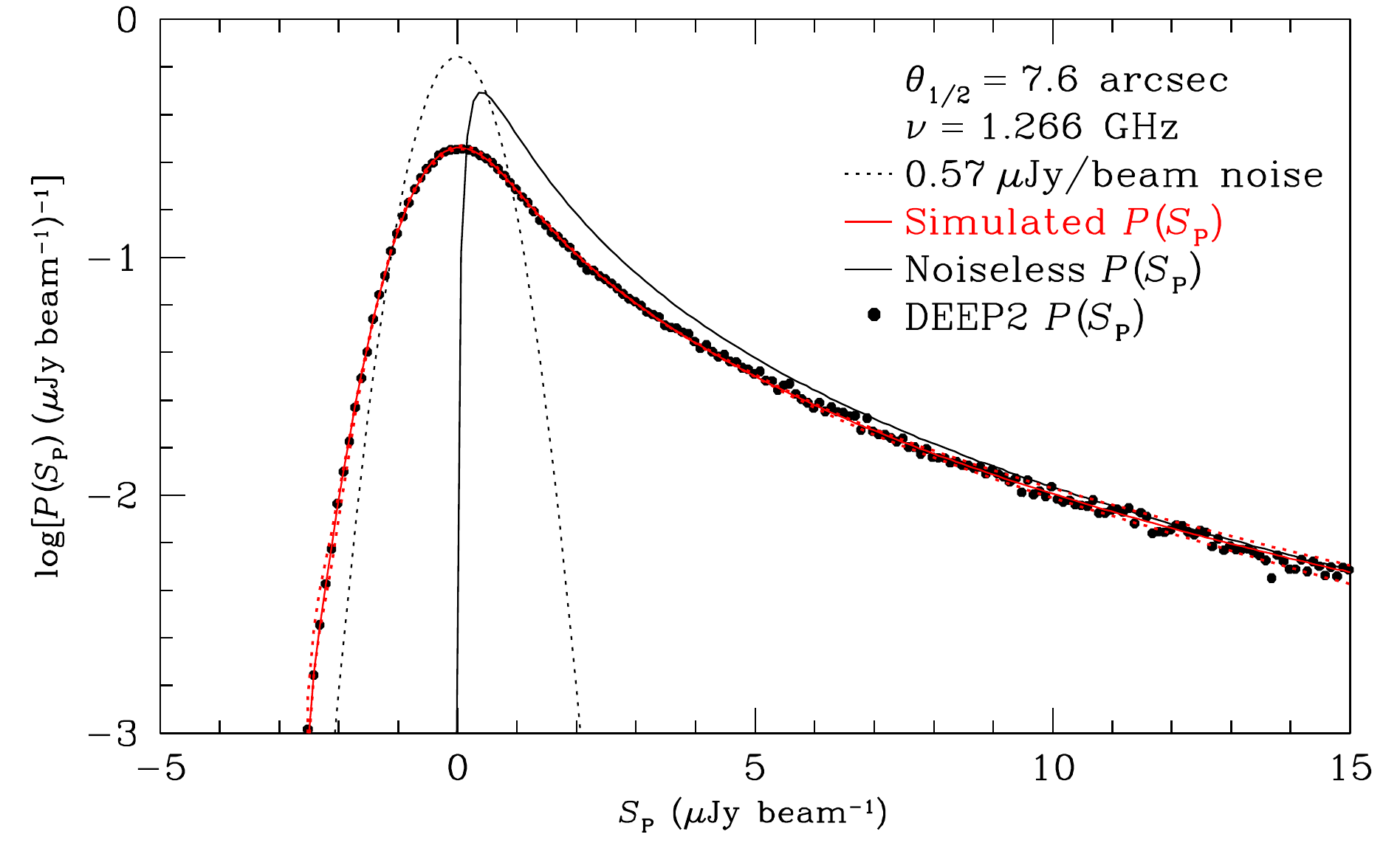}
    \caption{The normalized 1.266\,GHz $P(S_\mathrm{p}) = P(D)$
      distribution extracted from the central $r = 500\arcsec$ circle
      in the $\theta_{1/2} = 7\,\farcs 6$ resolution DEEP2 image corrected for
      primary beam attenuation is shown by the heavy black dots
      representing bins of width $\Delta S_\mathrm{p} = 0.1
      \,\mu\mathrm{Jy\,beam}^{-1}$.  It is quite smooth because it is
      based on $ 2.40 \times 10^4$ independent samples. The dotted
      parabola on this semilogarithmic plot represents the
      $\sigma_\mathrm{n} = 0.57\,\mu{\rm Jy\,beam^{-1}}$ Gaussian
      noise distribution 
inside the $P(D)$ circle.
      The red curve is the mean of 1000 simulated
      $P(D)$ distributions based on the best-fit source counts
      specified by Equation \ref{eq:bestcounts}, and the red dotted
      curves bound the range that includes 2/3 of those simulated
      $P(D)$ distributions.  The black curve shows the best-fit noiseless $P(D)$ distribution.    
These new fits are significantly more accurate than those shown in
\citet[fig.~13]{mauch20}, which were based on a power-law approximation to the
source counts.
    \label{fig:pd}}
\end{figure*}

The peak flux density $S_\mathrm{p}$ at any point in an image is the
sum of contributions from noise-free source confusion and image noise.
Confusion and noise are independent, so the observed $P(D)$
distribution is the convolution of the confusion and noise
distributions.  The noise amplitude distribution in an
aperture-synthesis image is easy to deconvolve because it is extremely
stable, Gaussian, and uniform across the image prior to correction for
primary-beam attenuation \citep{condon12}, unlike the noise in a
single-dish image, which usually varies significantly with position and time
during an observation.  Consequently we were able to measure the DEEP2
rms noise
and confirm its Gaussian amplitude distribution with very small
uncertainties.  The noise distribution is narrower than the noiseless $P(D)$
distribution in the very sensitive DEEP2 image, so we could deconvolve
the Gaussian noise distribution from the observed $P(D)$ distribution
to calculate the desired noiseless $P(D)$ distribution with
unprecedented accuracy and sensitivity.

Following the same procedure used in \citet{mauch20},
we extracted the $P(D)$ distribution from the circle of radius $r =
500''$ covering solid angle $\Omega = 1.85 \times
10^{-5}\,\mathrm{sr}$ centered on the SNR-weighted 1.266 GHz DEEP2
image corrected for primary-beam attenuation and shown in figure 11 of
\citet{mauch20}.  The $P(D)$ circle is small enough ($2r \ll
\Theta_{1/2}$) that the mean primary-beam attenuation is 0.98 inside
the circle and 0.96 at the edge, so its rms noise after correction for
primary-beam attenuation is only $\sigma_\mathrm{n} = (0.56 \pm
0.01\,\mu \mathrm{Jy\,beam}^{-1}) / 0.98 = 0.57 \pm 0.01
\,\mu\mathrm{Jy\,beam}^{-1}$.  The $P(D)$ circle is still large enough
to cover $N_\mathrm{b} = 1.20 \times 10^4$ restoring beam solid angles
$\Omega_\mathrm{b} = \pi \theta_{1/2}^2 / (4 \ln 2) = 1.54 \times
10^{-9} \,\mathrm{sr}$.  The solid angle of the \emph{square} of the
restoring beam attenuation pattern determines the number of
independent samples per unit solid angle of sky \citep{condon12}.  For
a Gaussian restoring beam, the solid angle of the beam squared is half
the beam solid angle, so the observed DEEP2 $P(D)$ distribution shown
by the large black points in Figure~\ref{fig:pd} actually contains
$2N_\mathrm{b} = 2.40 \times 10^4$ statistically independent samples.
The observed $P(D)$ distribution is the convolution of the noiseless
sky $P(D)$ distribution (black curve) with the $\sigma_\mathrm{n} =
0.57 \pm 0.01 \,\mu\mathrm{Jy\,beam}^{-1}$ Gaussian noise distribution
accurately represented by the parabolic dotted curve in the
semi-logarithmic Figure~\ref{fig:pd}.

The 1.266\,GHz DEEP2 $P(D)$ distribution shown in Figure~\ref{fig:pd}
is $4\times$ as sensitive to point sources with $\alpha \approx -0.7$
as the most sensitive published 3\,GHz $P(D)$ distribution
\citep{condon12}.  Such sources are $1.83 \times$ stronger at
1.266\,GHz than at 3\,GHz, so the rms noise $\sigma_\mathrm{n} = 1.255
\,\mu\mathrm{Jy\,beam}^{-1}$ of the 3\,GHz $P(D)$ distribution is
equivalent to $\sigma_\mathrm{n}= 2.30 \,\mu\mathrm{Jy\,beam}^{-1}$ at
1.266\,GHz.  The peak of the DEEP2 noise distribution is higher than
the peak of the noiseless $P(D)$ distribution (Figure~\ref{fig:pd}),
indicating that DEEP2 is strongly confusion limited, while the peak of
the 3\,GHz noise distribution is only half as high as the peak of the
3\,GHz noiseless $P(D)$ distribution.  The DEEP2 $P(D)$ distribution
also has smaller statistical uncertainties because it includes $3.1
\times$ as many independent samples.  Finally, the DEEP2 $P(D)$
distribution was extracted from the very center of the image where the
primary beam attenuation $\geq 0.96$, so systematic errors caused by
antenna pointing fluctuations or primary-beam attenuation corrections
are negligible.
  
\subsection{$P(D)$ statistical counts of
  $0.25 \leq S(\mu\mathrm{Jy}) \leq 10$ sources}

The noiseless confusion $P(D)$ distribution is sensitive to the
differential counts $n(S)$ of sources more than $10 \times$ fainter
than the usual $5\sigma_\mathrm{n}$ detection limit for individual
sources.
In terms of the number $N(>S)$ of sources per steradian
stronger than $S$, the mean number of sources stronger than $S$ per
beam solid angle $\Omega_\mathrm{b}$ is $\mu =
[N(>S)\Omega_\mathrm{b}]$ and the Poisson probability that \emph{all}
sources in a beam are weaker than $S$ is $P_\mathrm{P} = \exp(-\mu)$.
The DEEP2 image $\Omega_\mathrm{b} \approx 1.54 \times
10^{-9}\,\mathrm{sr}$ and the best-fit source counts from
\citet{mauch20}
imply $P_\mathrm{P} \approx 0.4$ at $S =
0.25\,\mu\mathrm{Jy\,beam}^{-1}$.  The DEEP2 $P(D)$ distribution was
extracted from a solid angle containing $2.40 \times 10^4$ independent
samples of the sky, so changes in the source count down to $S =
0.25\,\mu{\rm Jy\,beam^{-1}}$ can be detected statistically from the
$10^4$ independent samples that contain only fainter sources if the
rms noise $\sigma_{\rm n} \lesssim \sigma_{\rm c}$, the noise
  due to source confusion.


We used the radio sky simulations described in Section
  \ref{sec:model} to constrain the source counts consistent with the
  observed $P(D)$ distribution. The DEEP2 $P(D)$ distribution is
smoothed by Gaussian noise
with rms $\sigma_\mathrm{n} = 0.57 \pm 0.01\,\mu\mathrm{Jy\,beam}^{-1}$
which degrades its sensitivity to significantly fainter sources.  To
estimate the sensitivity of DEEP2 to faint sources in the presence of
noise, we simulated noisy $P(D)$ distributions using a variety of
differential source counts below $S = 10\,\mu{\rm Jy}$.  Above
$S=10\,\mu{\rm Jy}$, we used the direct source counts from DEEP2 and
the NVSS presented in Sections \ref{sec:deep2count} and
\ref{sec:nvsscount}.  The simulation accepts brightness-weighted
differential source counts $S^2n(S)$ specified in bins of width
$\Delta \log(S) = 0.2$.  Directly binning counts $n(S)$ that vary
rapidly with $S$ can introduce a significant bias \citep{jau68}.  We
mitigated this bias by binning the quantity $S^2 n(S)$ which changes
little across a flux-density bin.

The source counts $S^2n(S)$ in the flux range
  $-8 < \log[S\mathrm{(Jy)}] < -4.9$
are well fit by a cubic polynomial.
To measure goodness-of-fit for each input source count
we defined a statistic that quadratically combines the reduced
$\chi_{\rm P(D)}^2$ from differences between the simulated and observed
$P(D)$ distributions for all $S_p < 15\,\mu{\rm Jy\,beam^{-1}}$ with
the $\chi_{\rm DC}^2$ of the differences between the simulated
counts and the direct counts of DEEP2 sources in bins centered on
$\log S = -4.9$ and $-4.7$:
\begin{equation}\label{eq:gof}
\chi^2 = \sqrt{\left(\chi_{P(D)}^2\right)^2 + \Big(\chi_{\rm DC}^2\Big)^2}.
\end{equation}

We optimized the parameters for the cubic polynomial by minimizing
Equation \ref{eq:gof}.  We inspected the residuals $\Delta
N/\sigma_N$, where $N$ is the number of independent samples per bin
and $\sigma_N$ is the Poisson error per bin associated with the
observed $P(D)$ distribution, of the resulting best-fits for the
presence of correlations as a function of brightness $D$.  The
existence of a signal akin to red-noise in our residuals would imply
our counts under- or over-estimate the counts of sources in specific
flux density ranges.

The 3rd-degree polynomial source count
\begin{equation}\label{eq:bestcounts}
  \begin{split}
    \log[S^2n(S)] = 2.718 & + 0.405(\log S + 5)\\
    & - 0.020(\log S + 5)^2\\ & + 0.019(\log S+5)^3~,
  \end{split}
\end{equation}
where $S$ is the 1.266\,GHz flux density in Jy, gave the simulated
$P(D)$ distribution (red curve in Figure~\ref{fig:pd}) best fitting
the observed distribution (black points) while maintaining continuity
in the transition from the $P(D)$ to direct counts at $S =
10\,\mu\mathrm{Jy}$.

We converted the $1.266$\,GHz flux densities and brightness-weighted
source counts in Equations~\ref{eq:bestcounts}, \ref{eq:bounds}, and
\ref{eq:boundup} to the common source-count frequency $\nu =
1.4\,\mathrm{GHz}$ for sources with spectral index $\alpha = -0.7$
using
\begin{equation}
\begin{split}
  \log (S_{1.4\,\mathrm{GHz}}) & =  \log(S_{1.266\,\mathrm{GHz}}) +
  \alpha \log \biggl( \frac {1.4}{1.266} \biggr) \\
  & \approx \log(S_{1.266\,\mathrm{GHz}}) - 0.0306
\end{split}
\end{equation}
and
\begin{equation}
\begin{split}
  \log [S^2 n(S)]_{1.4\,\mathrm{GHz}} & = \log[S^2 n(S)]_{1.266\,\mathrm{GHz}} +
  \alpha \log \biggl( \frac {1.4}{1.266} \biggr) \\
  & \approx \log[S^2 n(S)]_{1.266\,\mathrm{GHz}} - 0.0306~.
\end{split}
\end{equation}
We also calculated the commonly used static-Euclidean source counts from
the brightness-weighted source counts via
\begin{equation}
  \log[S^{5/2} n(S)] = 0.5 \log(S) + \log[S^2 n(S)]~.
\end{equation}
Our 1.4\,GHz differential source counts with both normalizations are plotted
---along with the discrete source counts calculated in Sections
\ref{sec:deep2count} and \ref{sec:nvsscount}--- in
Figure~\ref{fig:counts}. In the following subsections, we describe
  our accounting of the various biases and uncertainties in our
  statistical fit of the source counts.

\subsection{Zero-level offset}\label{sec:zerolevel}

Before calculating the value of this statistic for a given simulation,
we removed the brightness zero-point offset between the simulated and
observed $P(D)$ distributions.  Numerous faint radio sources produce a
smooth background which is invisible to MeerKAT and other correlation
interferometers lacking zero-spacing data.  Thus the brightness zero
level of our observed $P(D)$ distribution is unknown and must be
fitted out.  We minimized the zero-level offset by comparing the
observed and simulated $P(D)$ distributions shifted in steps of
$0.001\,\mu\mathrm{Jy\,beam}^{-1}$, this step size being smaller than
the rms noise divided by the square root of the number
$2.40\times10^4$ of independent samples in the DEEP2 $P(D)$ area.

\subsection{DEEP2 rms noise uncertainty}\label{sec:noise}

\begin{figure}
  \centering
  \includegraphics[trim={0cm 0cm 3cm 4cm},clip,width=0.5\textwidth]
                 {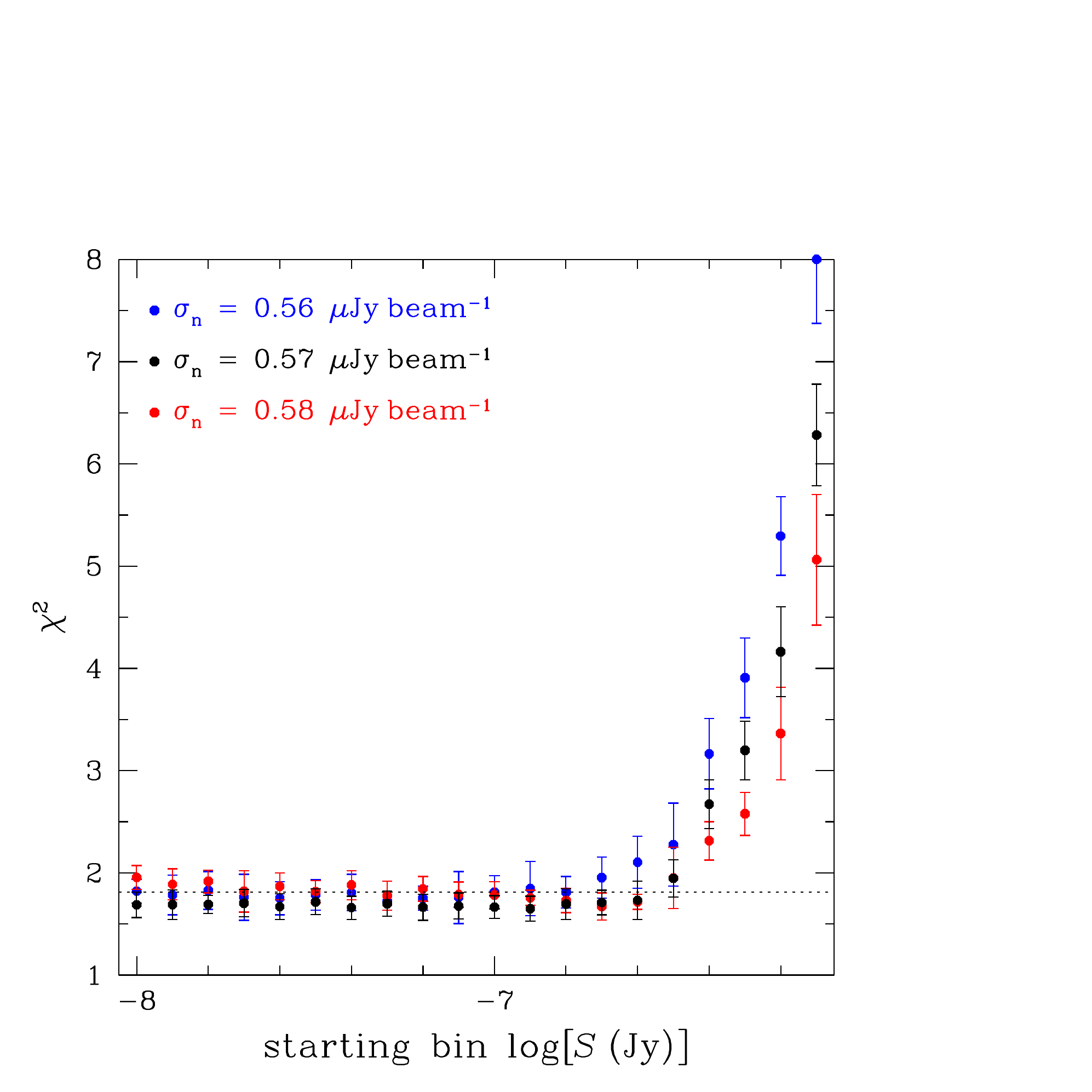}
 \caption{The reduced $\chi^2$ statistic from 1.266\,GHz DEEP2
   simulations is shown as a function of the starting source count bin
   increasing from the nominal $\log[ S\mathrm{(Jy)}] = -8$ to $ -6.1$.
     At $\log [S\mathrm{(Jy)}] = -6.5$, the $\chi^2$ statistic of the
       $P(D)$ distribution from DEEP2 simulations with
       $\sigma_\mathrm{n} = 0.57\,\mu{\rm Jy}$ averaged over the $r =
       500''$ $P(D)$ circle (black points) exceeds the value of the
       mean minimum $\chi^2$ plus $1\sigma$ (black dotted line). This
       indicates that we are sensitive to changes in the source count
       down to $\log [S\mathrm{(Jy)}] = -6.6$. The $\chi^2$ values for
         DEEP2 simulations with $\pm 1\sigma$ in rms noise are shown
         in blue and red for $0.56$ and $0.58\,\mu{\rm
           Jy\,beam^{-1}}$, respectively. \label{fig:lowerlimit}}
\end{figure}


Although the simulation includes sources as
faint as $S = 0.01 \,\mu\mathrm{Jy}$, the DEEP2 image is not sensitive
to the counts of such faint sources.  The biggest cause of uncertainty
in our sub-$\mu$Jy source counts is the $\pm 0.01\,\mu
\mathrm{Jy\,beam}^{-1}$ uncertainty in the DEEP2 rms noise.  To
estimate the flux density of the faintest sources we can usefully
count, we set the rms noise to $\sigma_\mathrm{n} =
0.57\,\mu\mathrm{Jy\,beam}^{-1}$ and iteratively removed the lowest
flux-density bin before using that sub-sample of bins to produce the
DEEP2 simulation.  After repeating this process for a total of 50
trials, we found that removing the source count bin at $\log
[S\mathrm{(Jy)}] = -6.6$ increases the simulation $\chi^2$ to more
than one-sigma above the mean minimum $\chi^2$, as shown by the the
black points above the dotted line in Figure \ref{fig:lowerlimit}.
The same process was repeated for for $\sigma_\mathrm{n} =
0.56\,\mu\mathrm{Jy\,beam}^{-1}$ (blue points) and $0.58 \,\mu{\rm
  Jy\,beam^{-1}}$ (red points).  The results are consistent with a
count sensitivity limit $\log [S\mathrm{(Jy)}] \approx -6.6$ or $S
\approx 0.25\,\mu \mathrm{Jy}$ in the presence of $\sigma_\mathrm{n}
= 0.57 \pm 0.01\,\mu\mathrm{Jy\,beam}^{-1}$ noise.

To determine the sensitivity of our best-fitting $P(D)$ distribution
above $S = 0.25\,\mu\mathrm{Jy}$ to small changes in the rms noise, we
ran 1000 simulations of the DEEP2 $P(D)$ distribution using the counts
given by Equation \ref{eq:bestcounts} and varying the
$\sigma_\mathrm{n} = 0.57 \,\mu\mathrm{Jy\,beam}^{-1}$ noise by adding
values drawn randomly from a Gaussian distribution of rms width
$0.01\,\mu{\rm Jy\,beam^{-1}}$.  The range of $P(D)$ containing 68\%
of these simulations best fitting (according to Equation \ref{eq:gof})
the average of all 1000 simulations defines the $\pm \sigma$
uncertainty region of our model $P(D)$. Figure \ref{fig:pd} includes
dotted red lines showing this narrow uncertainty region, which is
easily visible only in the $S_\mathrm{p} >
10\,\mu\mathrm{Jy\,beam}^{-1}$ tail of the distribution.

\subsection{Estimating the source-count uncertainty}\label{sigmas}

To estimate the $\pm \sigma$ source-count errors resulting from the
above $P(D)$ distribution range, we ran 500 simulations with the
following variations: (1) the noise was drawn randomly from Gaussian
distributions with mean $\sigma_\mathrm{n} = 0.57\,\mu\mathrm
{Jy\,beam^{-1}}$ and scatter $0.01 \,\mu\mathrm{Jy\,beam}^{-1}$; (2)
the input source counts were modeled with a fourth-degree polynomial
to allow for the rapidly growing count uncertainty at the lowest flux
densities caused by noise as well as by a possible ``new'' population
of very faint radio sources; and (3) the coefficients of the
fourth-degree polynomials were drawn randomly from gaussian
distributions centered on the best-fitting values given in Equation
\ref{eq:bestcounts} with an rms of 0.1 (the unknown fourth-degree
coefficient was initially centered on zero).

Each combination of coefficients and rms noise was repeated an
additional six times to determine the effects of noise on the
goodness-of-fit.  We considered a set of coefficients to be in
agreement with the DEEP2 $P(D)$ distribution if at least five of the
total seven simulations fell within the $\pm\sigma$ uncertainty region
determined from the original, un-altered 1000 simulations.  The subset
of the 500 coefficient-varying simulations that satisfied this
criterion define the statistical uncertainty of the measured source
counts.  Then we added quadratically a 3\% count uncertainty to absorb
possible 3\% systematic flux-density calibration errors.  In the flux
density range $-6.6 < \log [S(\rm Jy)] < 5$, the 1-$\sigma$ lower
limit of the 1.266\,GHz source-count error region is
\begin{equation}{\label{eq:bounds}}
\begin{split}
  \log[S^2n(S)] = 2.677 &+ 0.489(\log S+5)\\
  &+0.077(\log S+5)^2\\&+0.061(\log S+5)^3\\&-0.058(\log S+5)^4
\end{split}
\end{equation}
and the 1-$\sigma$ upper limit is
\begin{equation}{\label{eq:boundup}}
\begin{split}
  \log[S^2n(S)] = 2.768 & +0.367 (\log S+5)\\
  & -0.076 (\log S+5)^2\\& -0.009(\log S+5)^3\\& +0.023(\log S+5)^4~.
\end{split}
\end{equation}

\section{The DEEP2 Source Catalog and Direct Counts}
\label{sec:deep2cat}

We used the attenuation-corrected ``sky'' image to search for discrete
sources. The ``effective frequency'' of the wideband SNR-weighted DEEP2
image for sources with median spectral index $\langle \alpha \rangle
\approx -0.7$ is $\nu_\mathrm{e} = 1.266$\,GHz \citep{mauch20}. Even after
correction for primary-beam attenuation, the DEEP2 image is strongly
confusion limited with rms noise $\sigma_\mathrm{n} < 1.12\,\mu
\mathrm{Jy\,beam}^{-1}$ everywhere inside the primary beam half-power
circle.  Consequently our catalog brightness sensitivity
limit $S_\mathrm{p}(1.266\,\mathrm{GHz}) = 10
\,\mu\mathrm{Jy\,beam}^{-1} > 9 \sigma_\mathrm{n}$ is uniform over the
whole primary half-power circle,
unlike the variable sensitivity limit of a deep source catalog
extracted from an image still attenuated by the primary beam.  Nearly
all $\mu$Jy radio sources are unresolved by the $\theta_{1/2} =
7\,\farcs 6$ DEEP2 restoring beam, so the DEEP2 catalog should
be nearly complete for sources with total flux densities just above
$S(1.266\,\mathrm{GHz}) = 10\,\mu\mathrm{Jy}$.  The relatively
large DEEP2 restoring beam is actually advantageous because
incompleteness corrections for partially resolved sources can be large
and uncertain when the beam size is not much larger than the median
source size \citep{morrison10,owen18}.

\subsection{The DEEP2 Component Catalog}

We applied the \emph{Obit} \citep{cotton08} source-finding task FndSou
to the DEEP2 sky image inside the DEEP2 primary beam half-power circle.
FndSou searches for islands of contiguous pixels and decomposes each
island into elliptical Gaussian components as faint as $S_\mathrm{p} =
10\,\mu\mathrm{Jy\,beam}^{-1}$.  Most radio sources with
$S(3\,\mathrm{GHz}) \gtrsim 5\,\mu\mathrm{Jy}$ (equivalent to $S
\gtrsim 9\,\mu\mathrm{Jy}$ at 1.266\,GHz for $\langle \alpha \rangle =
-0.7$) have angular diameters $\phi \lesssim 0\,\farcs 66$
\citep{cotton18} and would be completely unresolved
in the DEEP2 image.  This point-source
approximation is supported by the qualitative similarity of our
point-source simulation and the actual DEEP2 image shown in Figure
\ref{fig:contours}.  A small fraction of the DEEP2 sources stronger
than $\sim 100\,\mu\mathrm{Jy}$ are
clearly resolved jets or
lobes driven by unresolved central AGNs, and they
can be represented by combining multiple components as described in
Section~\ref{sec:extendedsources}.

The sky density of sources reaches one per 25 restoring beam solid
angles at $S(1.266\,\mathrm{GHz}) \approx 17\,\mu\mathrm{Jy}$
\citep{mauch20}, so a significant fraction of our $S \gtrsim 10
\,\mu\mathrm{Jy}$ components partially overlap, and our catalog
accuracy, completeness, and reliability are limited more by confusion
than by noise.  To optimize the DEEP2 component catalog and understand
its limitations, we used FndSou to extract catalogs of components from
simulated images and compared those catalogs with the simulation input
source lists.  We compared catalogs in which the fitted elliptical
Gaussians were allowed to vary in width to ``point source'' catalogs
in which they were
not and found
that forcing point-source fits generally gave better matches to the
``true'' simulation input catalogs.  Therefore we forced point-source
fits to make the DEEP2 component catalog, and we later combined
components as needed to represent multicomponent extended radio
sources.

In a few crowded regions, FndSou reported spurious faint components
very close to much stronger sources.  To decide which components to
reject from our catalog, we generalized the original Rayleigh
criterion for resolving two equal point sources observed with an Airy
pattern PSF: the peak of one lies on or outside the first zero of the
second, which ensures that the total response has a minimum between
them.

\begin{figure}[!ht]
  \centering
  \includegraphics[trim = {3.5cm 9.5cm 4.5cm 9cm},clip, width = 0.49\textwidth]
    {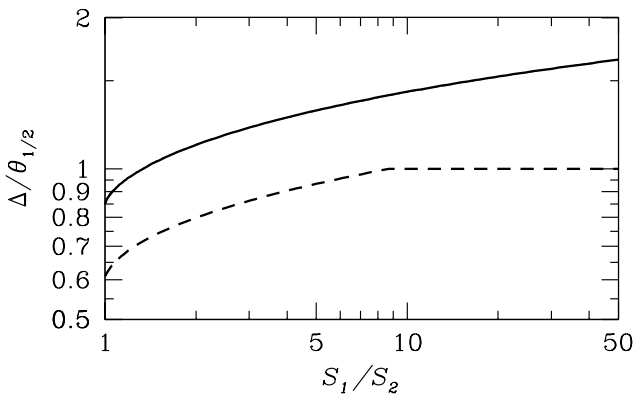}
    \caption{The continuous curve shows the calculated minimum
      separation $\Delta$ in Gaussian beamwidths $\theta_{1/2}$ needed
      to produce a minimum between two point sources as a function of
      their flux-density ratio $S_1/S_2$.  The dashed curve shows the
      empirical minimum separation for reliable DEEP2 components
      stronger than $10\,\mu\mathrm{Jy}$.
      \label{fig:minsep}}
\end{figure}

The total image response $R$ at position $x$ between unequal
components $S_1$ at $x_1 = 0$ and $S_2 < S_1$ at $x_2 = \Delta$ to a
Gaussian PSF with FWHM $\theta_{1/2}$ is
\begin{equation}\label{eqn:rmin}
  R = S_1 \exp\biggl( - \frac {4 \ln 2}{\theta_{1/2}^2} x^2 \biggr)
  + S_2 \exp \biggl[- \frac{4 \ln 2}{\theta_{1/2}^2} (x - \Delta)^2\biggr]~.
\end{equation}
For $R$ to have a minimum between the components, $dR/dx = 0$ for some
$0 < x < \Delta$.  The continuous curve in Figure~\ref{fig:minsep}
shows the required component separation $\Delta / \theta_{1/2}$ as a
function of the flux-density ratio $S_1 / S_2$.
All  DEEP2 catalog components stronger than $S > 10 \,\mu\mathrm{Jy}$
have SNR$>9$, so the requirement in
Equation~\ref{eqn:rmin} is stricter than necessary.  By comparing ten
catalogs of components extracted from simulated images with the ``true''
input components used to generate the simulated images, we found that faint
components near stronger components are reliable if they satisfy the
weaker criterion
\begin{eqnarray}\label{eqn:minsepemp}
 \frac{\Delta}{\theta_{1/2}} \geq 0.574 +0.357[\log(S_1/S_2) +
   0.01]^{1/2} + \qquad \hspace{0.8cm} \nonumber \\
 0.082 \log(S_1/S_2) \hspace{1.9cm} \mathrm{if~} (S_1 / S_2) < 9
 \qquad \nonumber \\
 \frac{\Delta}{\theta_{1/2}} \geq 1 \hspace{4.25cm} \mathrm{if~} (S_1
 / S_2) \geq 9 \qquad
  \end{eqnarray}
shown by the dashed curve in Figure~\ref{fig:minsep}.  We rejected the
334 probably spurious DEEP2 components ($<2$\% of the total) failing
to satisfy Equation~\ref{eqn:minsepemp}.

To estimate the effects of confusion and noise on the completeness,
reliability, positions, and flux densities of the surviving 17,350
DEEP2 components, we ran ten independent simulations of the DEEP2
field out to the  half-power circle of the primary beam using input
source counts consistent with the differential source counts in Table
\ref{tab:directcounts} and the 1.4\,GHz statistical count $S^2 n(S) =
1.07 \times 10^{-5} S^{-0.48} \,\mathrm{Jy\,sr}^{-1}$ of fainter
sources from \citet{mauch20}.  For each simulated image, we used
FndSou to find all components stronger than $10\,\mu{\rm Jy}$ and
rejected the components that did not satisfy the resolution criterion
in Equation~\ref{eqn:minsepemp}.  The positions and flux
densities of the resulting ten catalogs were compared with the
``true'' simulation input positions and flux densities of all
simulated sources stronger than $5\, \mu{\rm Jy}$.

We matched a simulated source to a cataloged component if (1) its
position was within $\theta_{1/2} / 2 = 3\,\farcs 8$ of the cataloged
position and (2) the catalog-to-true flux ratio satisfied $0.5 \leq
S_{\rm cat}/S_{\rm true}\leq 2$.  If there were two or more matches,
only the strongest simulated source was matched with the cataloged
component.  If there were no simulated sources that satisfied these
criteria, the cataloged component was rejected as spurious.  The ten
simulations yielded $\sim 1.6 \times 10^5$ matches.  Only $\sim 0.5$\%
of the cataloged components had no simulated-source counterpart, for
a catalog reliability $\approx 99.5$\%.

FndSou measures intensities relative to the image zero level.  The
DEEP2 interferometric image is insensitive to the smooth background of
very faint radio sources.  Our simulations of the radio sky brightness
include such a background, so the average confusion $P(D)$
distribution from ten simulations of DEEP2 appears shifted by $\Delta
D = +0.28\,\mu\mathrm{Jy\,beam^{-1}}$ compared with the $P(D)$
distribution of the real DEEP2 image.  The final flux densities of
components in the DEEP2 source catalog were corrected for the
zero-level offset by subtracting $0.28\,\mu\mathrm{Jy\,beam}^{-1}$
from the peak flux densities reported by FndSou.

\subsection{DEEP2 Catalog Position Uncertainties}

\begin{figure}[!ht]
  \centering
  \includegraphics[trim = {0cm 1cm 3.5cm 9cm},clip, width = 0.49\textwidth]
    {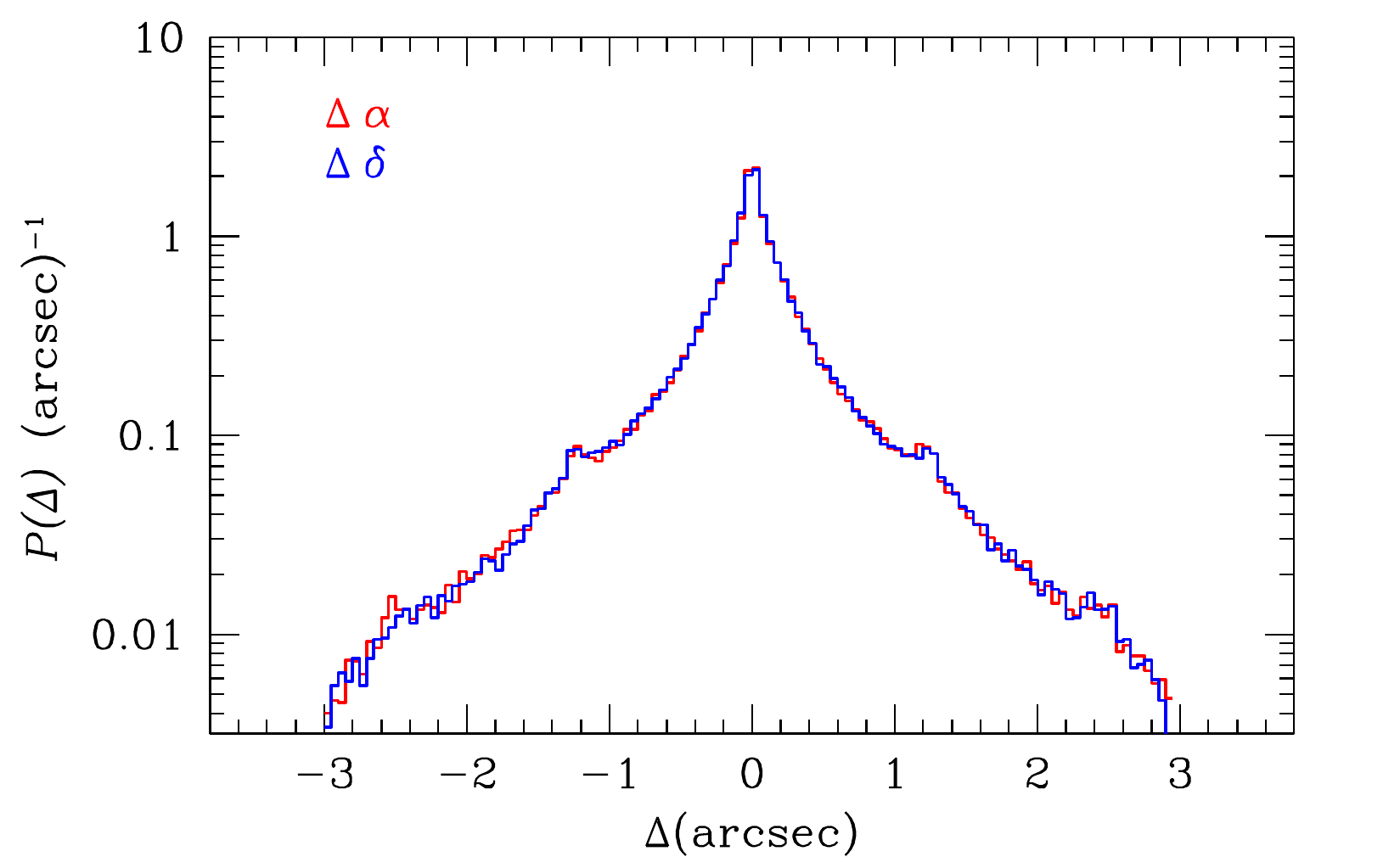}
  \caption{The distributions $P\, \mathrm{(arcsec)}^{-1}$ of
    differences in right ascension $\Delta \alpha$ (red curve) and
    declination $\Delta \delta$ (blue curve) between the cataloged and
    ``true'' simulation positions for all sources with $S \geq
    10\,\mu\mathrm{Jy}$ from ten simulated DEEP2 images.  The small
    peaks at integer multiples of $\Delta = 1\farcs 25$ are artifacts
    from measuring distances in simulated images composed of
    $1\farcs25$ pixels but do not affect the rms position
    errors.  \label{fig:radecdiff}}
\end{figure}

\begin{figure}[!ht]
  \centering
  \includegraphics[trim={0cm 5cm 4cm 6cm},clip,width=0.49\textwidth]
   {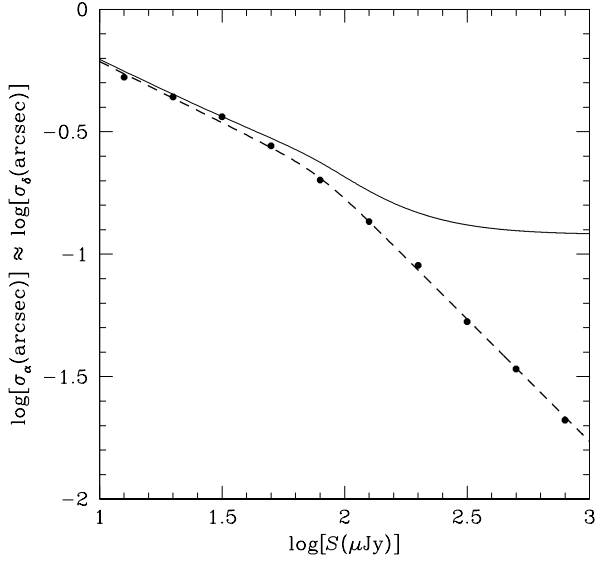}
  \caption{The simulation ``rms'' (defined as half the 68\% confidence interval)
    random position error in either right ascension $\alpha$
    and declination $\delta$ is shown as a function of flux density
    $S$ by the points.  The black dashed curve shows the best-fitting
    broken power-law from Equation \ref{eq:poserrors}.  The total
    DEEP2 ``rms'' position uncertainties estimated by the quadrature
    sum of Equation \ref{eq:poserrors} random errors and $0\,\farcs 12$
    systematic position errors are indicated by the solid
    curve.  \label{fig:pdelta}}
  \end{figure}

The random position errors of DEEP2 source components are dominated by
confusion errors whose non-Gaussian distributions are difficult to calculate
analytically, so we estimated the random position errors from
the differences $\Delta \alpha$, $\Delta \delta$ between the cataloged
and ``true'' input positions of source components in our ten
simulations.  The normalized probability distributions $P(\Delta)$ are
shown separately for right ascension $\alpha$ (red histogram) and
declination $\delta$ (blue histogram) in Figure \ref{fig:radecdiff}.
The distributions of $\Delta \alpha$ and $\Delta \delta$ are
indistinguishable, as expected for a circular PSF.  Also as expected,
the distributions of random positions errors are symmetrical about
$\Delta = 0$ and have long non-Gaussian tails\mdash a Gaussian distribution
would look like a parabola in the semilogarithmic
Figure~\ref{fig:radecdiff}. The formal rms width of $P(\Delta)$ is
dominated by these tails and is not a stable measure of the position
error distribution.  In a Gaussian distribution with rms $\sigma$,
68\% of the sources would lie within the range $-\sigma < \Delta <
+\sigma$, so we used the range of actual position offsets
$\Delta\alpha \approx \Delta\delta$ and defined the ``rms'' position
errors $\sigma_\alpha$ and $\sigma_\delta$ such that 68\% of the
components lie in the range $-\sigma_\alpha < \Delta\alpha <
+\sigma_\alpha$ or $-\sigma_\delta < \Delta\delta < +\sigma_\delta$.
The DEEP2 random position errors vary with component flux density $S$.
For bins of width 0.2 in $\log(S)$ centered on $\log[ S (\mu{\rm Jy})]
= 1.1,\, 1.3, \,\ldots,\, 2.9$, we determined the distributions of
position offsets $\Delta \alpha$ and $\Delta \delta$ in the ten
simulations.  Figure \ref{fig:pdelta} shows $\sigma_\alpha \approx
\sigma_\delta$ as a function of $\log(S)$.

The relationship between this error limit and flux density is well
described by a broken power-law of the form
\begin{equation}
  \label{eq:poserrors}
  \frac{\Delta \alpha}{\mathrm{arcsec}} =
  \frac{\Delta \delta}{\mathrm{arcsec}} =
  C \left[ \left(\frac{S_*}{S}\right)^{R/2} +
  \left(\frac{S_*}{S}\right)^{R}\right]^{1/R}~,
  \end{equation}
where the parameter $R$ controls the sharpness of the break at $S =
S_*$.  A nonlinear least squares fit to Equation \ref{eq:poserrors}
yields $R=-11.97$, $C = 0.219$, and $S_* = 78.6\,\mu{\rm Jy}$.  As the
catalogs were made for simulated images, there are no systematic
position errors included in Equation \ref{eq:poserrors}.  The dashed
black curve in Figure \ref{fig:pdelta} shows the random rms errors
$\sigma_\alpha \approx \sigma_\delta$ as a function of component flux density.
The slope of the dashed curves in Figures~\ref{fig:pdelta} and
  \ref{fig:fracfluxerrors} changes from $-1$ to $-0.5$ below $S_*$
  because the DEEP2 image is strongly limited by confusion, the
  source-count slope changes by $\Delta \gamma \approx -1$ near $S =
  S_*$, and the rms confusion from weaker sources is proportional to
  $S^{(3-\gamma)/2}$ \citep[see eq.~20 in][]{condon12}.  No such break
  occurs in noise-limited images.

To estimate the DEEP2 systematic position uncertainties and offsets,
we selected the 268 strong components with calculated random errors
$\sigma_\alpha = \sigma_\delta \leq 0\,\farcs05$ and used the
NASA/IPAC Infrared Science Archive (IRSA) to find eight
identifications with Gaia DR2 sources whose position errors are much
smaller than $0\,\farcs05$.  Their DEEP2 minus Gaia offsets have rms
$\sigma_\alpha = \sigma_\delta = 0\,\farcs12 \pm 0\,\farcs 04$, an
insignificant mean offset $+0\,\farcs03 \pm 0\,\farcs04$ in right
ascension, and a $3\sigma$ significant mean declination offset
$+0\,\farcs12 \pm 0\,\farcs04$.  We therefore added $-0\,\farcs12$ to
our fitted DEEP2 declinations and added the $0\,\farcs 12$ systematic
position errors to the random errors in quadrature to get the total
DEEP2 position error shown by the continuous curve in
Figure~\ref{fig:pdelta}.  The total position errors reported in
the final component catalog (Table
\ref{tab:cat}) reflect this quadrature sum and the corrected
declinations.

\begin{figure}
  \centering
  \includegraphics[trim={0cm 5cm 4cm 4cm}, clip, width=0.49\textwidth]
   {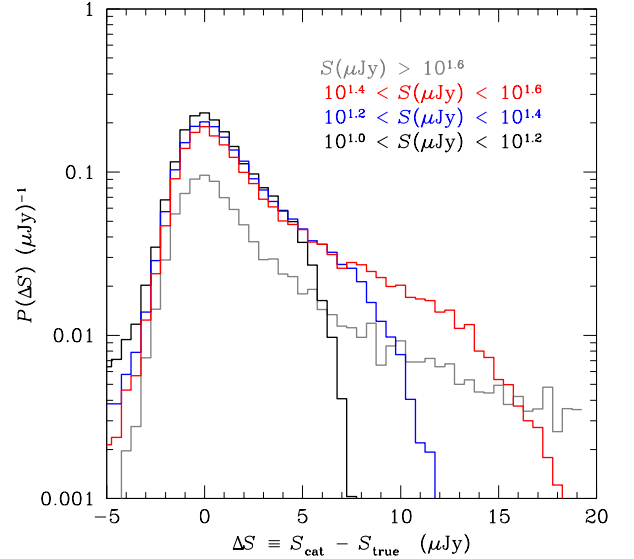}
   \caption{ The differences $\Delta S = S_\mathrm{cat} -
     S_\mathrm{true}$ between the cataloged and ``true'' simulation
     flux density are shown for four catalog flux-density
     ranges: $10\,\mu{\rm Jy} < S < 10^{1.2} \sim 16\,\mu{\rm Jy}$
     (black), $10^{1.2}\,\mu {\rm Jy} < S < 10^{1.4} \sim 25\,\mu{\rm
       Jy}$ (blue), $10^{1.4}\,\mu{\rm Jy} < S < 10^{1.6} \sim
     40\,\mu{\rm Jy}$ (red), and $S > 10^{1.6}\,\mu{\rm Jy}$
     (gray).  \label{fig:fluxerrors}}
  \end{figure}

\begin{figure}
  \centering
  \includegraphics[trim={0cm 5cm 4cm 4.5cm},clip, width=0.49\textwidth]
    {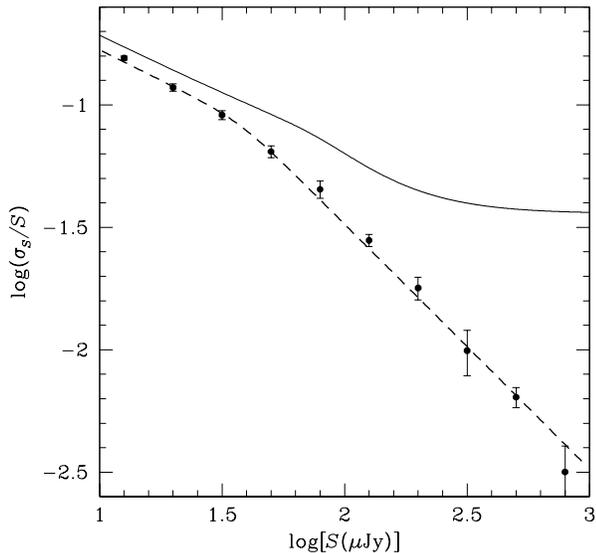}
  \caption{The random fractional flux density errors per catalog flux
    density bin was calculated as the narrowest range of the 
    distribution containing 68\% of the simulated sources (black points). The
    dashed line shows a broken power-law fit to these calculated
    errors.  The total component error $\sigma_S / S)$ from Equation
    \ref{eq:fracflux} is shown as the solid black curve.
    \label{fig:fracfluxerrors}}
\end{figure}

We compared the flux densities calculated from the forced point-source
fits of cataloged components in the ten simulated images with their
``true'' input flux densities.  The distribution of these differences
$\Delta S = S_\mathrm{cat} - S_\mathrm{true}$ is shown in Figure
\ref{fig:fluxerrors} for four flux density ranges: $10\,\mu{\rm Jy} <
S < 10^{1.2} \sim 16\,\mu{\rm Jy}$, $10^{1.2}\,\mu {\rm Jy} < S <
10^{1.4} \sim 25\,\mu{\rm Jy}$, $10^{1.4}\,\mu{\rm Jy} < S < 10^{1.6}
\sim 40\,\mu{\rm Jy}$, and $S > 10^{1.6}\,\mu{\rm Jy}$.  The
flux-density error distributions all peak near $\Delta S = 0\,\mu{\rm
  Jy}$ but have positive tails that grow with flux density because
stronger components are able to obscure stronger confusing components.

The fractional flux density errors $\sigma_S/S$ were calculated for
all ten catalogs of the simulated images in bins of width 0.2 dex
centered on $\log [S\,(\mu{\rm Jy})] = 1.1, 1.3, \ldots, 2.9$ and are
shown in Figure \ref{fig:fracfluxerrors}.  In the ideal case of
uncorrelated Gaussian noise, high signal-to-noise $S/\sigma_S$, and a
circular Gaussian beam of FWHM $\theta_{1/2}$, Equation 21
of \cite{condon97} gives
\begin{equation}
  \frac {\sigma_S}{S}= \sqrt{8 \ln 2}\left(\frac{\sigma_{\alpha}}{\theta_{1/2}}\right) =
  \sqrt{8\ln 2}\left(\frac{\sigma_{\delta}}{\theta_{1/2}}\right),
\end{equation}
so for either $\alpha$ or $\delta$,
\begin{eqnarray}
  \log\Biggl( \frac{\sigma_S}{S}\Biggr) &=&
  \log(8\ln2)/2 +\log(\sigma_{\alpha})
  -\log(\theta_{1/2}) \nonumber \\
  &=& \log(\sigma_{\alpha}) - 0.51,
\end{eqnarray}
for $\theta_{1/2} = 7\,\farcs 6$.  This gives conservative
flux-density fitting errors.  To them we add in quadrature a 2\%
uncertainty for telescope pointing errors and primary attenuation
uncertainty inside the primary beam half-power circle plus a 3\% for
the absolute flux-density uncertainty of the gain calibrator
PKS~B1934$-$638 \citep{mauch20} to get the total fractional
uncertainty
\begin{equation}
  \label{eq:fracflux}
  \frac{\sigma_S}{S} =
  \Biggl( 8\ln 2 \frac{\sigma_{\alpha}^2}{\theta_{1/2}^2} + 0.036^2 \Biggr)^{1/2}~.
  \end{equation}
The fractional flux density errors calculated from Equation
\ref{eq:fracflux} are shown by the solid black curve in Figure
\ref{fig:fracfluxerrors}.  This method yields more conservative error
estimates than directly fitting the measured flux differences from the
simulated images with a broken power law for flux densities $\log S <
2.5$ when these measured differences are added in quadrature with the
cumulative calibration uncertainties (shown as the dotted line in
Figure \ref{fig:fracfluxerrors}).

The ten simulated images were corrected for primary beam attenuation
before the catalog was created, so the noise contribution increases
distance $r$ from
the pointing center as
\begin{equation}
  \label{eq:noise}
  \sigma_\mathrm{n}(r) = \frac {0.56 \,\mu\mathrm{Jy\,beam}^{-1}} {a(r)},
\end{equation}
where 
\begin{equation}
  \label{eq:attenuation}
a(r) = \exp\left(-4\ln 2 \frac{r^2}{\Theta_{1/2}^2}\right),
\end{equation}
is the primary beam attenuation.  The
simulation placed sources in the sky randomly but uniformly, so we
subtracted the average noise variance within the half-power circle
$\langle \sigma_\mathrm{n}\rangle = 0.824\,\mu{\rm Jy \ beam}^{-1}$ from the
total average flux-density variance calculated from Equation
\ref{eq:fracflux}.  We added back the distance-dependent rms noise
variance to get a better estimate of errors on the individual source
flux densities at various distances from the pointing center.


\begin{deluxetable} {cccc}
  \tabletypesize{\footnotesize}
  \tablecaption{The 1.266 GHz DEEP2 Component Catalog   \label{tab:cat}}
  \tablehead{
    \colhead{Right Ascension} & \colhead{Declination} & \colhead{\llap{$S$}(1.266\,GHz)}  &
    \colhead{\llap{G}rou\rlap{p}}\\
    \colhead{(J2000)}                    & \colhead{(J2000)}     &  \colhead{($\mu$Jy)} &
    \colhead{\llap{C}od\rlap{e}}
  }
  \startdata
\llap{0}4:08:34.781 $\pm$ 0.17\rlap{0} & $-$79:51:43.08 $\pm$ 0.45 & \llap{2}0.0 $\pm$  2\rlap{.1}  & \nodata \\ 
\llap{0}4:08:34.897 $\pm$ 0.08\rlap{9} & $-$80:20:26.14 $\pm$ 0.22 & \llap{8}7.2 $\pm$  5\rlap{.3}  & \llap{G}0\rlap{9} \\
\llap{0}4:08:34.956 $\pm$ 0.16\rlap{6} & $-$79:35:02.48 $\pm$ 0.45 & \llap{2}0.0 $\pm$  2\rlap{.2}  & \nodata \\ 
\llap{0}4:08:34.986 $\pm$ 0.16\rlap{0} & $-$80:24:19.47 $\pm$ 0.40 & \llap{2}5.5 $\pm$  2\rlap{.5}  & \nodata \\ 
\llap{0}4:08:35.066 $\pm$ 0.05\rlap{1} & $-$79:47:01.95 $\pm$ 0.13 & \llap{27}8.0 $\pm$ 1\rlap{0.5}  & \nodata \\ 
  \enddata
  \tablecomments{Table~\ref{tab:cat} is published in its entirety in
    machine-readable format.  A portion is shown here for guidance
    regarding its form and content. The quoted uncertainties are
    similar to rms errors in that they encompass 68\% of the sources
    but are insensitive to the long tails of confusion-limited error
    distributions.  There are 35 multicomponent sources labeled by
    their component group numbers G01 through G35, as described in
    Section~\ref{sec:extendedsources}.}
\end{deluxetable}

\subsection{Multicomponent Sources}
\label{sec:extendedsources}

\begin{figure}
  \centering
  \includegraphics[trim={0.4cm 1cm 0cm 0.5cm},clip, width=0.47\textwidth]
    {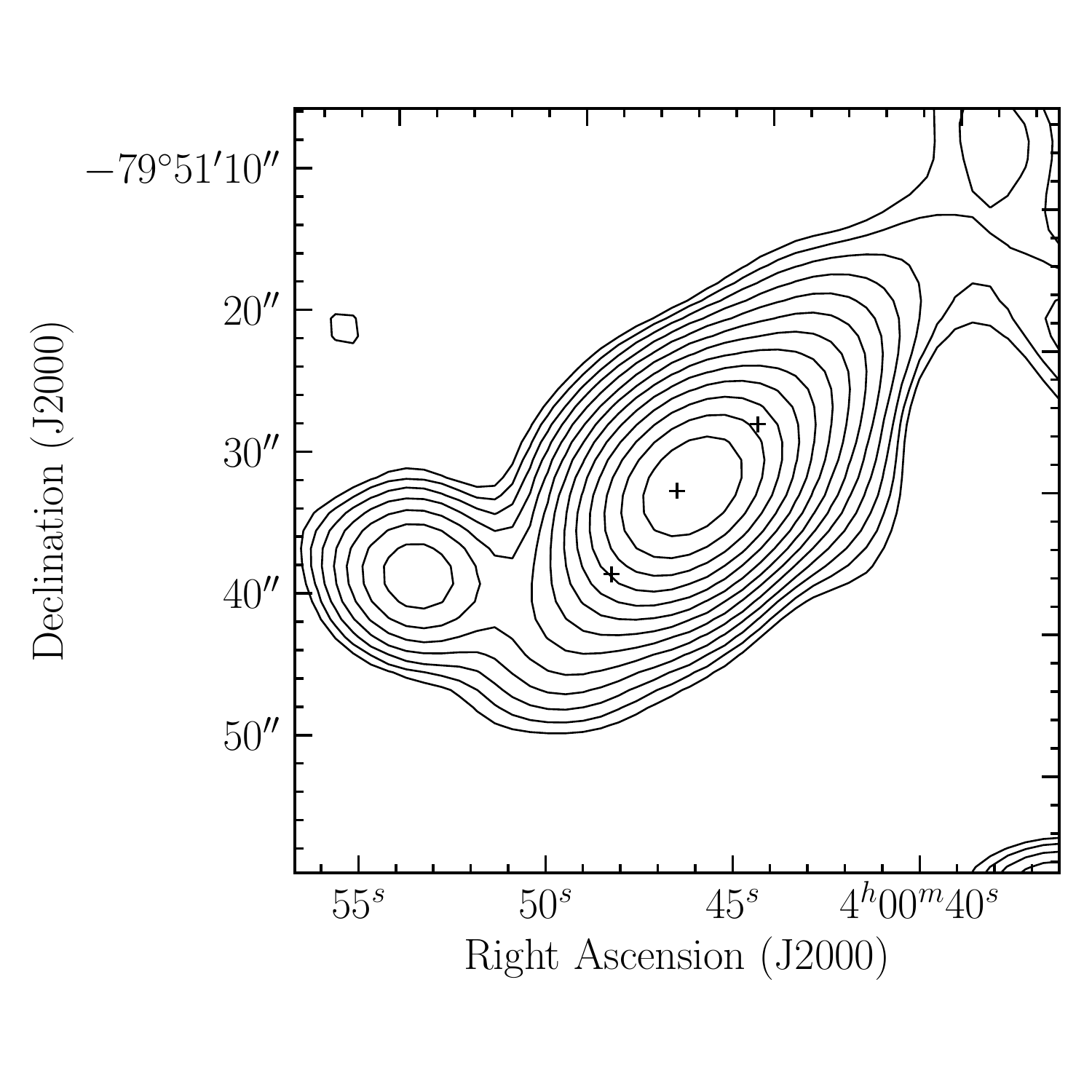}
  \caption{Multicomponent source G01.  Contour levels $\pm
    5\,\mu\mathrm{Jy\,beam}^{-1} \times 2^0, 2^{1/2}, 2^{1}, \dots$
    are plotted (negative contours, where present, shown as dashed
    lines).  Crosses mark the three components comprising this
    source. The complete figure set (35 images) is available in the
    online journal.\label{fig:multi}}
\end{figure}

We visually inspected the DEEP2 image and found 35 groups of
components that appear to comprise multicomponent radio sources.  We
labeled these components by their group numbers G01 to G35.  For an
extended source well approximated by a collection of Gaussian
components, we summed the individual component flux densities to
determine the source flux density.  For a source containing diffuse
emission regions, we estimated the flux density of such regions by
directly summing over the pixel brightness distribution.
Table~\ref{tab:groups} lists the 35 multicomponent sources, the number
of components in each source group, our best estimate of the source
core position, and the source flux density.   Figure~\ref{fig:multi}
shows the contour map of multicomponent source G01 with crosses
marking the positions of its three components.  Similar contour maps
of all multicomponent sources appear in the Appendix.

\begin{deluxetable}{crccr}[!ht]
\centering
\tablecaption{DEEP2 Multicomponent sources
\label{tab:groups}}  
\tablehead{
    \multicolumn{2}{c}{Group} & 
    \colhead{Right Ascension} & \colhead{Declination} & \colhead{$S$} \\
    \colhead{Code} & \colhead{$N$} &
    \colhead{(J2000)} & \colhead{(J2000)} & \colhead{($\mu$Jy)}
}
\startdata
G01  &    3  &   04:00:47.04   &  $-$79:51:31.4  &      817  \\
G02  &    3  &   04:01:31.81   &  $-$79:59:08.6  &      159  \\
G03  &    8  &   04:03:54.19   &  $-$80:08:49.1  &      757  \\
G04  &    4  &   04:04:05.81   &  $-$79:58:56.2  &      709  \\
G05  &    3  &   04:04:14.84   &  $-$79:56:21.5  &    13536  \\
G06  &    3  &   04:06:15.50   &  $-$80:10:57.5  &     2742  \\
G07  &    3  &   04:06:26.05   &  $-$79:38:01.0  &      168  \\
G08  &    7  &   04:06:27.83   &  $-$80:18:48.0  &    19200  \\
G09  &   14  &   04:08:42.38   &  $-$80:20:40.9  &      885  \\
G10  &   12  &   04:08:47.70   &  $-$80:24:02.4  &     1569  \\
G11  &    5  &   04:11:32.60   &  $-$79:48:41.3  &      766  \\
G12  &    3  &   04:11:38.97   &  $-$79:48:17.5  &     2867  \\
G13  &   11  &   04:11:59.21   &  $-$80:14:54.8  &     1422  \\
G14  &    7  &   04:12:16.93   &  $-$79:46:33.2  &     6772  \\
G15  &    5  &   04:12:32.00   &  $-$79:34:36.4  &      509  \\
G16  &    8  &   04:13:24.93   &  $-$79:49:21.2  &     5455  \\
G17  &    9  &   04:13:41.22   &  $-$79:46:34.9  &     6680  \\
G18  &    4  &   04:13:58.03   &  $-$79:42:19.2  &     1369  \\
G19  &   18  &   04:14:17.93   &  $-$80:11:38.4  &     5455  \\
G20  &    5  &   04:14:58.85   &  $-$80:29:08.4  &     1261  \\
G21  &    3  &   04:16:10.11   &  $-$80:03:31.9  &      169  \\
G22  &    6  &   04:16:23.34   &  $-$80:20:54.5  &     4699  \\
G23  &    3  &   04:16:47.09   &  $-$79:48:50.3  &    54268  \\
G24  &    5  &   04:16:58.32   &  $-$79:54:46.2  &     1953  \\
G25  &    9  &   04:17:02.19   &  $-$80:12:33.9  &     6115  \\
G26  &    3  &   04:17:06.86   &  $-$79:51:28.6  &     3682  \\
G27  &    7  &   04:18:58.15   &  $-$79:51:23.5  &     1603  \\
G28  &   10  &   04:19:10.77   &  $-$80:30:32.4  &     3051  \\
G29  &    4  &   04:20:03.10   &  $-$80:27:11.3  &     2769  \\
G30  &   14  &   04:22:05.41   &  $-$80:03:30.1  &    10882  \\
G31  &    5  &   04:23:19.81   &  $-$79:51:10.6  &    11408  \\
G32  &    8  &   04:25:02.63   &  $-$80:14:15.9  &    26271  \\
G33  &    6  &   04:25:18.38   &  $-$79:52:22.3  &    15983  \\
G34  &    9  &   04:25:51.5\hphantom{0}    &  $-$79:54:38\hphantom{.0}  &      731  \\
G35  &    8  &   04:26:07.75   &  $-$80:09:23.7  &      766  \\
  \enddata
\end{deluxetable}

\figsetstart
\figsetnum{11}
\figsettitle{Multicomponent Source Contour Maps}

\figsetgrpstart
\figsetgrpnum{11.1}
\figsetgrptitle{G01:\quad 04:00:47 ~$-$79:51:31}
\figsetplot{GRP1_1.PDF}
\figsetgrpnote{Contour levels 
      $\pm 5\,\mu\mathrm{Jy\,beam}^{-1} \times 2^0, 2^{1/2}, 2^{1}, \dots$ are plotted.
      Crosses mark the source group components.}
\figsetgrpend

\figsetgrpstart
\figsetgrpnum{11.2}
\figsetgrptitle{G02:\quad 04:01:31.81  ~$-$79:59:08.5}
\figsetplot{GRP2_1.PDF}
\figsetgrpnote{Contour levels 
      $\pm 5\,\mu\mathrm{Jy\,beam}^{-1} \times 2^0, 2^{1/2}, 2^{1}, \dots$ are plotted.
      Crosses mark the source group components.}
\figsetgrpend

\figsetgrpstart
\figsetgrpnum{11.3}
\figsetgrptitle{G03:\quad 04:03:54.9 ~$-$80:08:49.0}
\figsetplot{GRP3_1.PDF}
\figsetgrpnote{Contour levels 
      $\pm 5\,\mu\mathrm{Jy\,beam}^{-1} \times 2^0, 2^{1/2}, 2^{1}, \dots$ are plotted.
      Crosses mark the source group components.}
\figsetgrpend

\figsetgrpstart
\figsetgrpnum{11.4}
\figsetgrptitle{G04:\quad 04:04:05.81    ~$-$79:58:56.1}
\figsetplot{GRP4_1.PDF}
\figsetgrpnote{Contour levels 
      $\pm 5\,\mu\mathrm{Jy\,beam}^{-1} \times 2^0, 2^{1/2}, 2^{1}, \dots$ are plotted.
      Crosses mark the source group components.}
\figsetgrpend

\figsetgrpstart
\figsetgrpnum{11.5}
\figsetgrptitle{G05:\quad 04:04:14.843   ~$-$79:56:21.4}
\figsetplot{GRP5_1.PDF}
\figsetgrpnote{Contour levels 
      $\pm 5\,\mu\mathrm{Jy\,beam}^{-1} \times 2^0, 2^{1/2}, 2^{1}, \dots$ are plotted.
      Crosses mark the source group components.}
\figsetgrpend

\figsetgrpstart
\figsetgrpnum{11.6}
\figsetgrptitle{G06:\quad 04:06:15.60    ~$-$80:10:57.4}
\figsetplot{GRP6_1.PDF}
\figsetgrpnote{Contour levels 
      $\pm 5\,\mu\mathrm{Jy\,beam}^{-1} \times 2^0, 2^{1/2}, 2^{1}, \dots$ are plotted.
      Crosses mark the source group components.}
\figsetgrpend

\figsetgrpstart
\figsetgrpnum{11.7}
\figsetgrptitle{G07:\quad 04:06:26.05    ~$-$79:38:00.9}
\figsetplot{GRP7_1.PDF}
\figsetgrpnote{Contour levels 
      $\pm 5\,\mu\mathrm{Jy\,beam}^{-1} \times 2^0, 2^{1/2}, 2^{1}, \dots$ are plotted.
      Crosses mark the source group components.}
\figsetgrpend

\figsetgrpstart
\figsetgrpnum{11.8}
\figsetgrptitle{G08:\quad 04:06:27.83    ~$-$80:18:47.9}
\figsetplot{GRP8_1.PDF}
\figsetgrpnote{Contour levels 
      $\pm 5\,\mu\mathrm{Jy\,beam}^{-1} \times 2^0, 2^{1/2}, 2^{1}, \dots$ are plotted.
      Crosses mark the source group components.}
\figsetgrpend

\figsetgrpstart
\figsetgrpnum{11.9}
\figsetgrptitle{G09:\quad 04:08:42.38    ~$-$80:20:40.8}
\figsetplot{GRP9_1.PDF}
\figsetgrpnote{Contour levels 
      $\pm 5\,\mu\mathrm{Jy\,beam}^{-1} \times 2^0, 2^{1/2}, 2^{1}, \dots$ are plotted.
      Crosses mark the source group components.}
\figsetgrpend

\figsetgrpstart
\figsetgrpnum{11.10}
\figsetgrptitle{G10:\quad 04:08:47.70   ~$-$80:24:02.3}
\figsetplot{GRP10_1.PDF}
\figsetgrpnote{Contour levels 
      $\pm 5\,\mu\mathrm{Jy\,beam}^{-1} \times 2^0, 2^{1/2}, 2^{1}, \dots$ are plotted.
      Crosses mark the source group components.}
\figsetgrpend

\figsetgrpstart
\figsetgrpnum{11.11}
\figsetgrptitle{G11:\quad 04:11:32.60   ~$-$79:48:41.2}
\figsetplot{GRP11_1.PDF}
\figsetgrpnote{Contour levels 
      $\pm 5\,\mu\mathrm{Jy\,beam}^{-1} \times 2^0, 2^{1/2}, 2^{1}, \dots$ are plotted.
      Crosses mark the source group components.}
\figsetgrpend

\figsetgrpstart
\figsetgrpnum{11.12}
\figsetgrptitle{G12:\quad 04:11:38.97   ~$-$79:48:17.4}
\figsetplot{GRP12_1.PDF}
\figsetgrpnote{Contour levels 
      $\pm 5\,\mu\mathrm{Jy\,beam}^{-1} \times 2^0, 2^{1/2}, 2^{1}, \dots$ are plotted.
      Crosses mark the source group components.}
\figsetgrpend

\figsetgrpstart
\figsetgrpnum{11.13}
\figsetgrptitle{G13:\quad 04:11:59.21   ~$-$80:14:54.7}
\figsetplot{GRP13_1.PDF}
\figsetgrpnote{Contour levels 
      $\pm 5\,\mu\mathrm{Jy\,beam}^{-1} \times 2^0, 2^{1/2}, 2^{1}, \dots$ are plotted.
      Crosses mark the source group components.}
\figsetgrpend

\figsetgrpstart
\figsetgrpnum{11.14}
\figsetgrptitle{G14:\quad 04:12:16.93   ~$-$79:46:33.1}
\figsetplot{GRP14_1.PDF}
\figsetgrpnote{Contour levels 
      $\pm 5\,\mu\mathrm{Jy\,beam}^{-1} \times 2^0, 2^{1/2}, 2^{1}, \dots$ are plotted.
      Crosses mark the source group components.}
\figsetgrpend

\figsetgrpstart
\figsetgrpnum{11.15}
\figsetgrptitle{G15:\quad 04:12:32.00   ~$-$79:34:36.3}
\figsetplot{GRP15_1.PDF}
\figsetgrpnote{Contour levels 
      $\pm 5\,\mu\mathrm{Jy\,beam}^{-1} \times 2^0, 2^{1/2}, 2^{1}, \dots$ are plotted.
      Crosses mark the source group components.}
\figsetgrpend

\figsetgrpstart
\figsetgrpnum{11.16}
\figsetgrptitle{G16:\quad 04:13:24.93   ~$-$79:49:21.1}
\figsetplot{GRP16_1.PDF}
\figsetgrpnote{Contour levels 
      $\pm 5\,\mu\mathrm{Jy\,beam}^{-1} \times 2^0, 2^{1/2}, 2^{1}, \dots$ are plotted.
      Crosses mark the source group components.}
\figsetgrpend

\figsetgrpstart
\figsetgrpnum{11.17}
\figsetgrptitle{G17:\quad 04:13:41.22   ~$-$79:46:34.8}
\figsetplot{GRP17_1.PDF}
\figsetgrpnote{Contour levels 
      $\pm 5\,\mu\mathrm{Jy\,beam}^{-1} \times 2^0, 2^{1/2}, 2^{1}, \dots$ are plotted.
      Crosses mark the source group components.}
\figsetgrpend

\figsetgrpstart
\figsetgrpnum{11.18}
\figsetgrptitle{G18:\quad 04:13:58.03   ~$-$79:42:19.1}
\figsetplot{GRP18_1.PDF}
\figsetgrpnote{Contour levels 
      $\pm 5\,\mu\mathrm{Jy\,beam}^{-1} \times 2^0, 2^{1/2}, 2^{1}, \dots$ are plotted.
      Crosses mark the source group components.}
\figsetgrpend

\figsetgrpstart
\figsetgrpnum{11.19}
\figsetgrptitle{G19:\quad 04:14:17.93   ~$-$80:11:38.3}
\figsetplot{GRP19_1.PDF}
\figsetgrpnote{Contour levels 
      $\pm 5\,\mu\mathrm{Jy\,beam}^{-1} \times 2^0, 2^{1/2}, 2^{1}, \dots$ are plotted.
      Crosses mark the source group components.}
\figsetgrpend

\figsetgrpstart
\figsetgrpnum{11.20}
\figsetgrptitle{G20:\quad 04:14:58.85   ~$-$80:29:08.3}
\figsetplot{GRP20_1.PDF}
\figsetgrpnote{Contour levels 
      $\pm 5\,\mu\mathrm{Jy\,beam}^{-1} \times 2^0, 2^{1/2}, 2^{1}, \dots$ are plotted.
      Crosses mark the source group components.}
\figsetgrpend

\figsetgrpstart
\figsetgrpnum{11.21}
\figsetgrptitle{G21:\quad 04:16:10.11   ~$-$80:03:31.8}
\figsetplot{GRP21_1.PDF}
\figsetgrpnote{Contour levels 
      $\pm 5\,\mu\mathrm{Jy\,beam}^{-1} \times 2^0, 2^{1/2}, 2^{1}, \dots$ are plotted.
      Crosses mark the source group components.}
\figsetgrpend

\figsetgrpstart
\figsetgrpnum{11.22}
\figsetgrptitle{G22:\quad 04:16:23.34   ~$-$80:20:54.4}
\figsetplot{GRP22_1.PDF}
\figsetgrpnote{Contour levels 
      $\pm 5\,\mu\mathrm{Jy\,beam}^{-1} \times 2^0, 2^{1/2}, 2^{1}, \dots$ are plotted.
      Crosses mark the source group components.}
\figsetgrpend

\figsetgrpstart
\figsetgrpnum{11.23}
\figsetgrptitle{G23:\quad 04:16:47.09   ~$-$79:48:50.2}
\figsetplot{GRP23_1.PDF}
\figsetgrpnote{Contour levels 
      $\pm 5\,\mu\mathrm{Jy\,beam}^{-1} \times 2^0, 2^{1/2}, 2^{1}, \dots$ are plotted.
      Crosses mark the source group components.}
\figsetgrpend

\figsetgrpstart
\figsetgrpnum{11.24}
\figsetgrptitle{G24:\quad 04:16:58.32   ~$-$79:54:46.1}
\figsetplot{GRP24_1.PDF}
\figsetgrpnote{Contour levels 
      $\pm 5\,\mu\mathrm{Jy\,beam}^{-1} \times 2^0, 2^{1/2}, 2^{1}, \dots$ are plotted.
      Crosses mark the source group components.}
\figsetgrpend

\figsetgrpstart
\figsetgrpnum{11.25}
\figsetgrptitle{G25:\quad 04:17:02.19   ~$-$80:12:33.8}
\figsetplot{GRP25_1.PDF}
\figsetgrpnote{Contour levels 
      $\pm 5\,\mu\mathrm{Jy\,beam}^{-1} \times 2^0, 2^{1/2}, 2^{1}, \dots$ are plotted.
      Crosses mark the source group components.}
\figsetgrpend

\figsetgrpstart
\figsetgrpnum{11.26}
\figsetgrptitle{G26:\quad 04:17:06.86   ~$-$79:51:28.5}
\figsetplot{GRP26_1.PDF}
\figsetgrpnote{Contour levels 
      $\pm 5\,\mu\mathrm{Jy\,beam}^{-1} \times 2^0, 2^{1/2}, 2^{1}, \dots$ are plotted.
      Crosses mark the source group components.}
\figsetgrpend

\figsetgrpstart
\figsetgrpnum{11.27}
\figsetgrptitle{G27: \quad 04:18:58.15   ~$-$79:51:23.4}
\figsetplot{GRP27_1.PDF}
\figsetgrpnote{Contour levels 
      $\pm 5\,\mu\mathrm{Jy\,beam}^{-1} \times 2^0, 2^{1/2}, 2^{1}, \dots$ are plotted.
      Crosses mark the source group components.}
\figsetgrpend

\figsetgrpstart
\figsetgrpnum{11.28}
\figsetgrptitle{G28:\quad 04:19:10.77   ~$-$80:30:32.3}
\figsetplot{GRP28_1.PDF}
\figsetgrpnote{Contour levels 
      $\pm 5\,\mu\mathrm{Jy\,beam}^{-1} \times 2^0, 2^{1/2}, 2^{1}, \dots$ are plotted.
      Crosses mark the source group components.}
\figsetgrpend

\figsetgrpstart
\figsetgrpnum{11.29}
\figsetgrptitle{G29:\quad 04:20:03.10   ~$-$80:27:11.2}
\figsetplot{GRP29_1.PDF}
\figsetgrpnote{Contour levels 
      $\pm 5\,\mu\mathrm{Jy\,beam}^{-1} \times 2^0, 2^{1/2}, 2^{1}, \dots$ are plotted.
      Crosses mark the source group components.}
\figsetgrpend

\figsetgrpstart
\figsetgrpnum{11.30}
\figsetgrptitle{G30:\quad 04:22:05.41 $-$80:03:30.0}
\figsetplot{GRP30_1.PDF}
\figsetgrpnote{Contour levels 
      $\pm 5\,\mu\mathrm{Jy\,beam}^{-1} \times 2^0, 2^{1/2}, 2^{1}, \dots$ are plotted.
      Crosses mark the source group components.}
\figsetgrpend

\figsetgrpstart
\figsetgrpnum{11.31}
\figsetgrptitle{G31:\quad 04:23:19.81   ~$-$79:51:10.5}
\figsetplot{GRP31_1.PDF}
\figsetgrpnote{Contour levels 
      $\pm 5\,\mu\mathrm{Jy\,beam}^{-1} \times 2^0, 2^{1/2}, 2^{1}, \dots$ are plotted.
      Crosses mark the source group components.}
\figsetgrpend

\figsetgrpstart
\figsetgrpnum{11.32}
\figsetgrptitle{G32:\quad 04:25:02.63   ~$-$80:14:15.8}
\figsetplot{GRP32_1.PDF}
\figsetgrpnote{Contour levels 
      $\pm 5\,\mu\mathrm{Jy\,beam}^{-1} \times 2^0, 2^{1/2}, 2^{1}, \dots$ are plotted.
      Crosses mark the source group components.}
\figsetgrpend

\figsetgrpstart
\figsetgrpnum{11.33}
\figsetgrptitle{G33:\quad 04:25:18.38   ~$-$79:52:22.2}
\figsetplot{GRP33_1.PDF}
\figsetgrpnote{Contour levels 
      $\pm 5\,\mu\mathrm{Jy\,beam}^{-1} \times 2^0, 2^{1/2}, 2^{1}, \dots$ are plotted.
      Crosses mark the source group components.}
\figsetgrpend

\figsetgrpstart
\figsetgrpnum{11.34}
\figsetgrptitle{G34:\quad 04:25:51.5    ~$-$79:54:38\hphantom{.0}}
\figsetplot{GRP34_1.PDF}
\figsetgrpnote{Contour levels 
      $\pm 5\,\mu\mathrm{Jy\,beam}^{-1} \times 2^0, 2^{1/2}, 2^{1}, \dots$ are plotted.
      Crosses mark the source group components.}
\figsetgrpend

\figsetgrpstart
\figsetgrpnum{11.35}
\figsetgrptitle{G35:\quad 04:26:07.75   ~$-$80:09:23.6}
\figsetplot{GRP35_1.PDF}
\figsetgrpnote{Contour levels 
      $\pm 5\,\mu\mathrm{Jy\,beam}^{-1} \times 2^0, 2^{1/2}, 2^{1}, \dots$ are plotted.
      Crosses mark the source group components.}
\figsetgrpend

\figsetend

\subsection{DEEP2 Direct Source Counts}\label{sec:deep2count}

We counted DEEP2 sources in bins of width 0.2 dex centered on
1.266\,GHz flux densities $\log[S({\rm Jy})] = -4.9, -4.7, \ldots,
-2.5$.  Component groups comprising an extended source were counted as
a single source whose flux density is the sum of its individual
component flux densities.  For the few extended sources with diffuse emission
regions, we estimated the flux densities of these regions by directly
summing over their pixel brightness distributions.

Sources near the catalog lower limit $S(1.266\,\mathrm{GHz}) =
10\,\mu{\rm Jy}$ may be biased up by confusion or biased down and
missed entirely.  We estimated the effects of confusion on the direct
source counts by comparing the measured counts in the ten simulated
images with their ``true'' input counts.  Their differences in each
flux-density bin were calculated individually for the ten simulations.
We added the mean differences $\Delta$ in $\log[S^2 n(S)]$ from the
simulations to the raw DEEP2 counts to yield more accurate counts of
radio sources with $-5.0 < \log[S\mathrm{(Jy)}] < -2.5$.

Table \ref{tab:directcounts} shows the $\nu = 1.266\,\mathrm{GHz}$
corrected counts based on the 17,350 DEEP2 components with $S >
10\,\mu\mathrm{Jy}$ inside the
half-power circle of the
primary beam.  For the 13 flux-density bins of width 0.2 in $\log(S)$,
Column 1 lists the bin center $\log[S\mathrm{(Jy)}]$ and column 2
lists the number $N_\mathrm{bin}$ of sources in the bin.  The
corrections $\Delta$ in column 3 were added to the values of $\log[S^2
  n(S) \,\mathrm{(Jy\,sr}^{-1}\mathrm{)]}$ in column 4.  Columns 5 and
6 are the rms positive and negative uncertainties in $\log[S^2 n(S)]$.
These uncertainties are the quadratic sum of the Poisson uncertainties
in samples of size $N_\mathrm{bin}$, the count correction
uncertainties which we conservatively estimate to be $\Delta / 2$, and
a 3\% overall flux-density scale uncertainty.

\begin{deluxetable}{c r c c}
\centering  \tablecaption{DEEP2 1.266 GHz direct source counts
\label{tab:directcounts}}
\tablehead{
  \colhead {$ \log [S\mathrm{(Jy)}]$} &
  \colhead {$N_{\rm bin}$} & \colhead {$\Delta$} &
  \colhead {$\log[S^2 n(S)\, \mathrm{(Jy~sr}^{-1}\mathrm{)}]$} 
} 
\startdata
$-$4.90  &   5504  & $+$0.054 &  2.730~  +0.028 $-$0.029  \\
$-$4.70  &   4460  & $+$0.020 &  2.801~  +0.017 $-$0.018  \\
$-$4.50  &   3010  & $-$0.022 &  2.788~  +0.018 $-$0.019  \\
$-$4.30  &   1998  & $-$0.027 &  2.802~  +0.020 $-$0.021  \\
$-$4.10  &   1053  & $-$0.034 &  2.716~  +0.025 $-$0.027  \\
$-$3.90  &    559  & $-$0.035 &  2.639~  +0.029 $-$0.031  \\
$-$3.70  &    289  & $-$0.028 &  2.563~  +0.031 $-$0.034  \\
$-$3.50  &    126  & $-$0.012 &  2.423~  +0.040 $-$0.044  \\
$-$3.30  &     72  & $-$0.011 &  2.366~  +0.050 $-$0.057  \\
$-$3.10  &     46  & $-$0.005 &  2.393~  +0.061 $-$0.072  \\
$-$2.90  &     24  & $-$0.006 &  2.305~  +0.082 $-$0.102  \\
$-$2.70  &     15  & $-$0.001 &  2.317~  +0.101 $-$0.132  \\
$-$2.50  &     18  & $-$0.009 &  2.587~  +0.093 $-$0.119
\enddata
\end{deluxetable}

\section{NVSS Source Counts} \label{sec:nvsscount}

The NVSS catalog reports flux densities rounded to the nearest
multiple of $0.1 \mathrm{~mJy}$.  For example, all NVSS components with
fitted flux densities ${2.45 \leq S \mathrm{(mJy)} < 2.55}$ are
listed as having $S = 2.5 \mathrm{~mJy}$.  We separated the NVSS
components into flux-density bins of nearly constant  width
0.2 in $\log(S)$ whose exact boundaries
$S_\mathrm{min}$ and $S_\mathrm{max}$ are midway between multiples
of 0.1~mJy.  Thus the lowest flux-density bin covers $2.45 \leq S
\mathrm{(mJy)} < 3.95$ and includes all NVSS components listed with $S
\mathrm{(mJy)} = 2.5,\, 2.6,\, 2.7,\, \dots, \, 3.9$.  The first 
column of Table~\ref{tab:nvsscounts} lists the bin centers, the
second lists the numbers $n_\mathrm{bin}$ of components in each
bin, and the third column shows the brightness-weighted counts
$\log[S^2 n(S)]$ at
$\log[S\mathrm{(Jy)}] = -2.5$, $-2.3$, \dots, $+1.3$.  The fourth
and fifth columns are the total rms uncertainties in $\log[S^2 n(S)]$.

Complex radio sources significantly more extended than the $45\arcsec$
FWHM NVSS restoring beam may be represented by two or more catalog
components, and such large multicomponent sources are more common at
flux densities $S \gtrsim 1 \mathrm{~Jy}$.  To estimate the fraction
of components comprising strong extended sources, we compared the NVSS
component catalog with the low-resolution 1.4~GHz \citet{bridle72}
catalog of 424 sources having $S \geq 1.7 \mathrm{~Jy}$ and equivalent
angular diameters $\phi \lesssim 10\arcmin$ in the area defined
by $-5\degr < \delta < +70\degr$, $\vert b \vert > 5\degr$.  We
combined NVSS components within $\sim 5\arcmin$ of each
\citet{bridle72} source, after excluding those that appeared to be
unrelated background sources.  In the six flux-density bins centered
on $\log[S\mathrm{(Jy)}] = +0.3$ through $+1.3$, grouping NVSS
components into sources changed the brightness-weighted source count
$\log[S^2 n(S) \mathrm{(Jy~sr}^{-1}\mathrm{)}]$ by $-0.013$, +0.013,
+0.109, +0.193, +0.133, and 0.000, respectively.

\begin{deluxetable}{c r c}
\centering  \tablecaption{NVSS 1.4 GHz source counts
\label{tab:nvsscounts}}
\tablehead{
  \colhead {$ \log [S\mathrm{(Jy)}]$} &
  \colhead {$N_{\rm bin}$} &
  \colhead {$\log[S^2 n(S)\, \mathrm{(Jy~sr}^{-1}\mathrm{)}]$} 
}
\startdata
 $ -2.50$ &  350531 & 2.475~ +0.030 $-$0.030 \\
 $ -2.30$ &  217350 & 2.498~ +0.019 $-$0.019 \\
 $ -2.10$ &  161525 & 2.598~ +0.015 $-$0.015 \\
 $ -1.90$ &  120302 & 2.676~ +0.014 $-$0.014 \\
 $ -1.70$ &   90072 & 2.743~ +0.013 $-$0.013 \\
 $ -1.50$ &   63706 & 2.793~ +0.013 $-$0.013 \\
 $ -1.30$ &   43803 & 2.831~ +0.013 $-$0.013 \\
 $ -1.10$ &   29035 & 2.849~ +0.013 $-$0.013 \\
 $ -0.90$ &   18094 & 2.844~ +0.013 $-$0.013 \\
 $ -0.70$ &   10891 & 2.822~ +0.013 $-$0.013 \\
 $ -0.50$ &    5950 & 2.757~ +0.014 $-$0.014 \\
 $ -0.30$ &    3089 & 2.670~ +0.015 $-$0.015 \\
 $ -0.10$ &    1477 & 2.551~ +0.017 $-$0.017 \\
 $ +0.10$ &     718 & 2.434~ +0.021 $-$0.021 \\
 $ +0.30$ &     303 & 2.247~ +0.028 $-$0.028 \\
 $ +0.50$ &     143 & 2.137~ +0.038 $-$0.038 \\
 $ +0.70$ &      51 & 1.995~ +0.080 $-$0.080 \\
 $ +0.90$ &      15 & 1.739~ +0.140 $-$0.140 \\
 $ +1.10$ &       6 & 1.523~ +0.214 $-$0.229 \\
 $ +1.30$ &       3 & 1.285~ +0.295 $-$0.341 
\enddata
\end{deluxetable}

 The differential source count $n(S)$ is a rapidly declining function
 of flux density, so simply counting the number of sources in each
 fairly wide flux-density bin throws away flux-density information and
 can bias the resulting estimate of $n(S)$.  If $n(S) \, dS$ is the
 number of sources per steradian with flux densities between $S$ and
 $S+dS$ and $\eta(S)\,d\ln(S)$ is the number per steradian with flux
 densities between $S$ and $S + d \ln(S)$, then $n(S) dS =
 \eta(S)\,d\ln S$ and $n(S) = \eta(S) / S$. We added the flux density
 of each source into its bin of logarithmic width $\Delta \approx
 \mathrm{dex}(0.2)$ to calculate the more nearly constant quantity
\begin{equation}
  S^2 n(S) = S \eta(S) =
  \Biggl[ \frac {1} {\Omega \ln (\Delta)} \Biggr]
  \displaystyle \sum_{i = 1}^{n_\mathrm{bin}} S_i
\end{equation}
directly.

Finally, counts in the faintest bins must be corrected for
population-law bias \citep{murdoch73}: faint sources outnumber strong
sources, so noise moves more faint sources into a bin than it moves
strong sources out. We used their Table 2 and the cumulative
source-count approximation $N(>S) \equiv \int_S^\infty n(S) \,dS
\propto S^{-1}$ for $S \gtrsim 2.5 \mathrm{~mJy}$ to calculate the
required corrections to $\log[S^2 n(S)]$.  They are $-0.030$,
$-0.012$, and $-0.004$ in bins centered on $\log[S\mathrm{(Jy)}] =
-2.5$, $-2.3$, and $-2.1$, respectively.

The rms statistical uncertainty in $S^2 n(S)$ for each bin with
$n_\mathrm{bin} \gg 1$ is
\begin{equation}
\sigma_\mathrm{stat} \approx \Biggl[\frac {1} {\Omega
    \ln(\Delta)} \Biggr] \Biggl(\displaystyle \sum_{i = 1}^{n_\mathrm{bin}}
S_i^2\Biggr)^{1/2}~.
\end{equation}
There are only $n_\mathrm{bin} = 6$ sources in the
${\log[S\mathrm{(Jy)}] = +1.1}$ bin and $n_\mathrm{bin} = 3$ sources
in the ${\log[S\mathrm{(Jy)}] = +1.3}$ bin, so we replaced their rms
statistical errors in $\log[ S^2 n(S)]$ by the \citet{gehrels86} 84\%
confidence-level errors $+0.203, -0.219$ and $+0.295, -0.341$,
respectively.  To these statistical uncertainties we added
quadratically the 3\% error in $S^2n(S)$ caused by the 3\% NVSS
flux-density scale uncertainty \citep{condon98} and systematic
uncertainties equaling half the corrections for component grouping
and population-law bias.

\begin{figure*}
  \includegraphics[width=\textwidth,trim={1cm 5.5cm 1cm 2.3cm},clip]{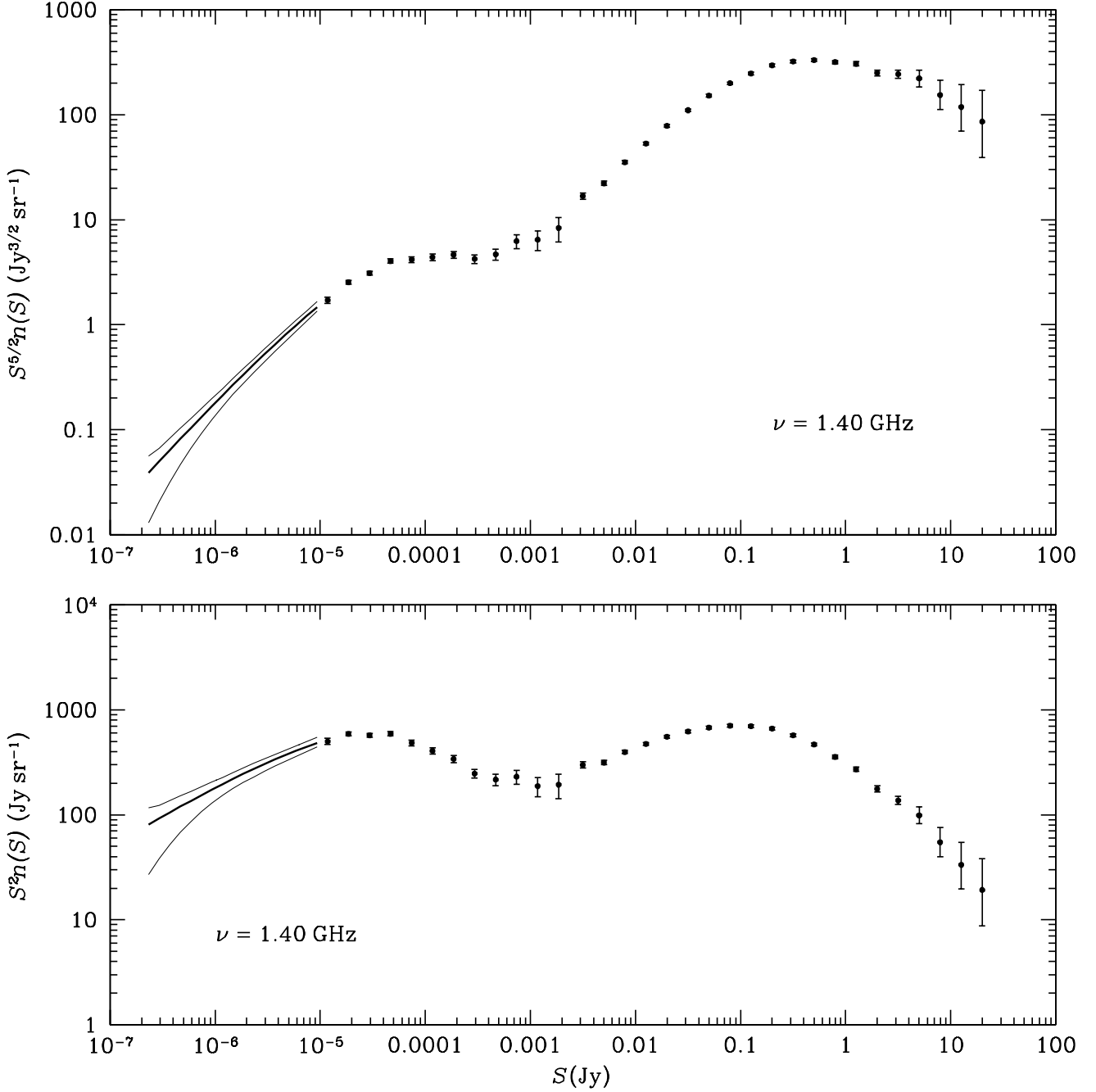}
  \caption{Our 1.4 GHz differential source counts between $0.25
    \,\mu\mathrm{Jy}$ and $25\,\mathrm{Jy}$ are shown with both the
    traditional static Euclidean normalization $S^{5/2}n(S)$ (top
    panel) and the brightness-weighted normalization $S^2 n(S)$
    (bottom panel).  The heavy curve and light $\pm 1\,\sigma$ error
    curves from $S = 2.5 \times 10^{-7}\,\mathrm{Jy}$ to
    $10^{-6}\,\mathrm{Jy}$ are statistical counts derived from the
    DEEP2 confusion $P(D)$ distribution (Section~\ref{sec:pofd}).  The
    black data points and their $\pm 1\,\sigma$ error bars show the
    1.4~GHz DEEP2 source counts between $S = 10^{-5} \, \mathrm{Jy}$
    and $S = 0.0025$\,Jy (Section~\ref{sec:deep2cat}) plus the NVSS
    counts (Section~\ref{sec:nvsscount}) at higher flux densities.
      \label{fig:counts}
      }
\end{figure*}

\section{Discussion}\label{sec:discussion}

Figure~\ref{fig:counts} shows our 1.4\,GHz differential source counts
with  traditional static-Euclidean weighting $S^{5/2} n(S)$
and with brightness weighting $S^2 n(S)$.  Counts from $S =
0.25\,\mu\mathrm{Jy}$ to $S = 10\,\mu\mathrm{Jy}$ were derived
statistically from the DEEP2 confusion $P(D)$ distribution extracted
from solid angle $\Omega =
0.061 \mathrm{\,deg}^2$.
Individual sources uniformly covering solid angle $\Omega =
1.04\mathrm{\,deg}^2$ between
$10\,\mu\mathrm{Jy}$ and $2.5\,\mu$Jy were counted directly, as were
NVSS sources above $2.5\,\mathrm{mJy}$ in solid angle $\Omega =
7.016\,\mathrm{sr}$ (0.56 of the sky). Together these counts span the
eight decades in flux density from $\log[S\mathrm{(Jy)}] = -6.6$ to
$\log[S\mathrm{(Jy)}] = +1.4$.  Their largest fractional uncertainties
are caused by the rms noise $\sigma_\mathrm{n} = 0.57 \pm 0.01
\,\mu\mathrm{Jy\,beam}^{-1}$ and finite resolution $\theta_{1/2} =
7\,\farcs6$ just above $S = 0.25\,\mu\mathrm{Jy}$, by statistical
fluctuations in the small numbers of sources in the
DEEP2 half-power circle between 
$0.5\,\mathrm{mJy}$ and $2.5\, \mathrm{mJy}$, and by cosmic variance in
the NVSS counts
above $S \approx 3\,\mathrm{Jy}$.

\begin{figure*}[!ht]
  \includegraphics[width=\textwidth,trim={1cm 9.5cm 1cm 6cm},clip]{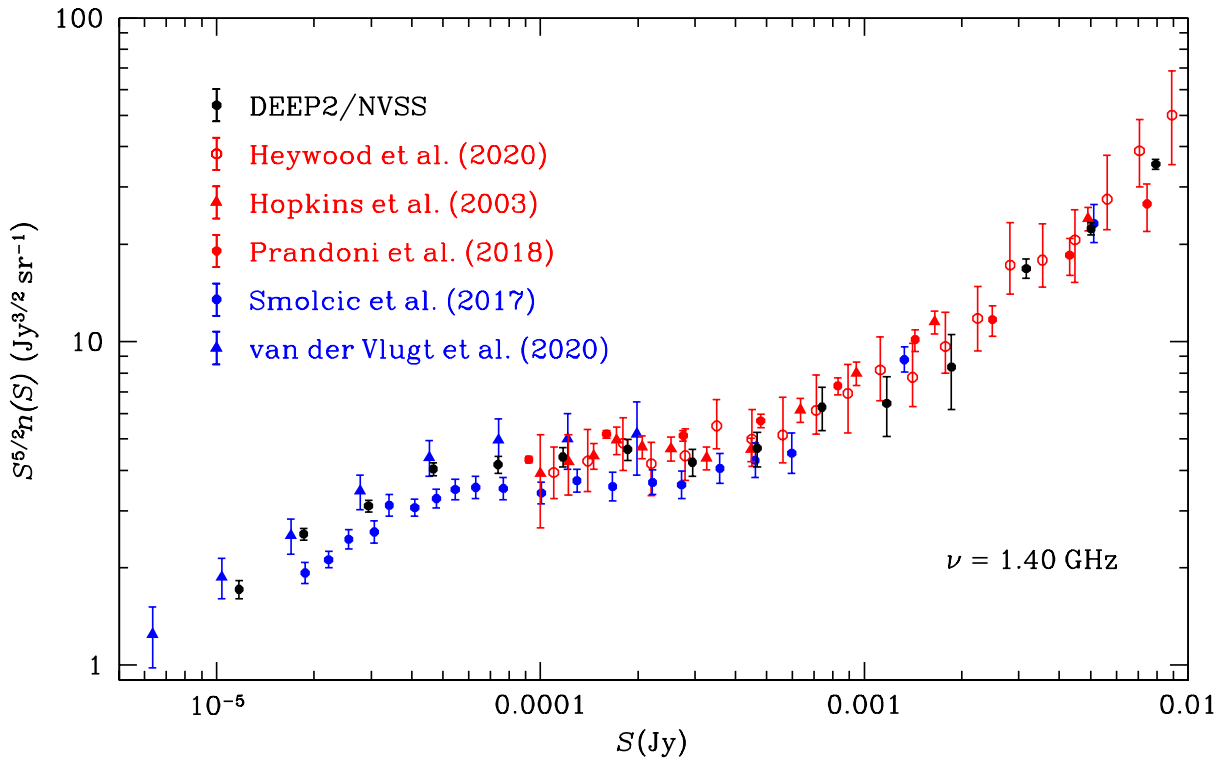}
  \caption{The 1.4 GHz differential source counts between $10
    \,\mu\mathrm{Jy}$ and $10\,\mathrm{mJy}$ are shown with the
    traditional static Euclidean normalization $S^{5/2}n(S)$. 
      The black data points show the DEEP2 source counts below
      $S = 0.0025$\,Jy (Section~\ref{sec:deep2cat}) and the NVSS counts
      (Section~\ref{sec:nvsscount} at higher flux densities.  The red data
      points are 1.4\,GHz counts from \citet{heywood20} (open circles),
      \citet{hopkins03} (solid triangles), and \citet{prandoni18} (filled
      circles).  The blue data points are based on the
      \citet{smolcic17} and the \citet{vandervlugt20}
      3~GHz counts converted to 1.4~GHz with a spectral
      index $\alpha = -0.7$.
      \label{fig:compare}
      }
\end{figure*}

Figure~\ref{fig:compare} compares our 1.4\,GHz direct counts (black
points) of sources fainter than 10\,mJy with those of
\citet{hopkins03} (red triangles), \citet{prandoni18} (filled red points),
\citet{heywood20} (open red  points), plus the \citet{smolcic17}
 (filled blue points) and \citet{vandervlugt20} (blue triangles) 3\,GHz
counts converted to 1.4\,GHz assuming the median spectral index is
$\langle \alpha \rangle = -0.7$.  The \citet{smolcic17} counts
have small ($\sigma \sim 10$\%) uncertainties because they are based
on the large (10,830 sources) noise-limited (median $\sigma_\mathrm{n}
= 2.3\,\mu\mathrm{Jy\,beam}$ at 3\,GHz) VLA-COSMOS catalog.  They are
$\sim 20\% \sim 2\sigma$ lower than most other counts and $\sim 30$\%
lower than the \citet{vandervlugt20} counts, possibly because
resolution corrections for the small ($\theta_{1/2} = 0\,\farcs75$)
VLA COSMOS beam were insufficient or the median spectral index of
$\mu$Jy sources is more negative than the assumed $-0.7$.
In any case, the agreement among all of these $\mu$Jy
source counts is much better than other counts have agreed in the
recent past \citep{heywood13}, suggesting that the large earlier
discrepancies were caused by observational and analysis errors, not by
surprisingly strong source clustering.

\subsection{Resolving the AGN and SFG backgrounds}\label{sec:knownsources}

The 1.4\,GHz brightness-weighted counts $S^2 n(S)$ shown in
Figure~\ref{fig:counts} have two broad peaks.  The peak  $S \sim
0.1\,\mathrm{Jy}$ is dominated by AGNs and the peak at $S \sim 3 \times
10^{-5}$ Jy by SFGs.  If the counts below $S = 0.25\,\mu\mathrm{Jy}$
do not exceed the extrapolation with slope $d \log [S^2 n(S)] / d
\log(S) = +0.5$, sources stronger $S = 0.25\,\mu\mathrm{Jy}$ resolve $
> 99$\% of the AGN contribution $\Delta T_\mathrm{b} \approx
0.06\,\mathrm{K}$ to the sky brightness temperature and $>96$\% of the
$\Delta T_\mathrm{b} \approx 0.04\,\mathrm{K}$ SFG contribution. Thus
most of the stars in the universe were formed in SFGs stronger than
$0.25\,\mu\mathrm{Jy}$.  For example, our fairly typical Galaxy currently has
1.4\,GHz spectral luminosity $L_\nu \approx 2.5 \times
10^{21}\,\mathrm{W\,Hz}^{-1}$.  With $10 \times$ luminosity evolution
\citep{madau14}, it would be a $1.2\,\mu\mathrm{Jy}$ source around
 ``cosmic noon'' at $z \sim 2$ and a $0.25\,\mu\mathrm{Jy}$ source
even at $z = 4$.  

\subsection{$P(D)$ limits on ``new'' source populations}

\begin{figure}
  \centering
  \includegraphics[trim={1.5cm 6cm .6cm 8cm},clip,width=0.5\textwidth]{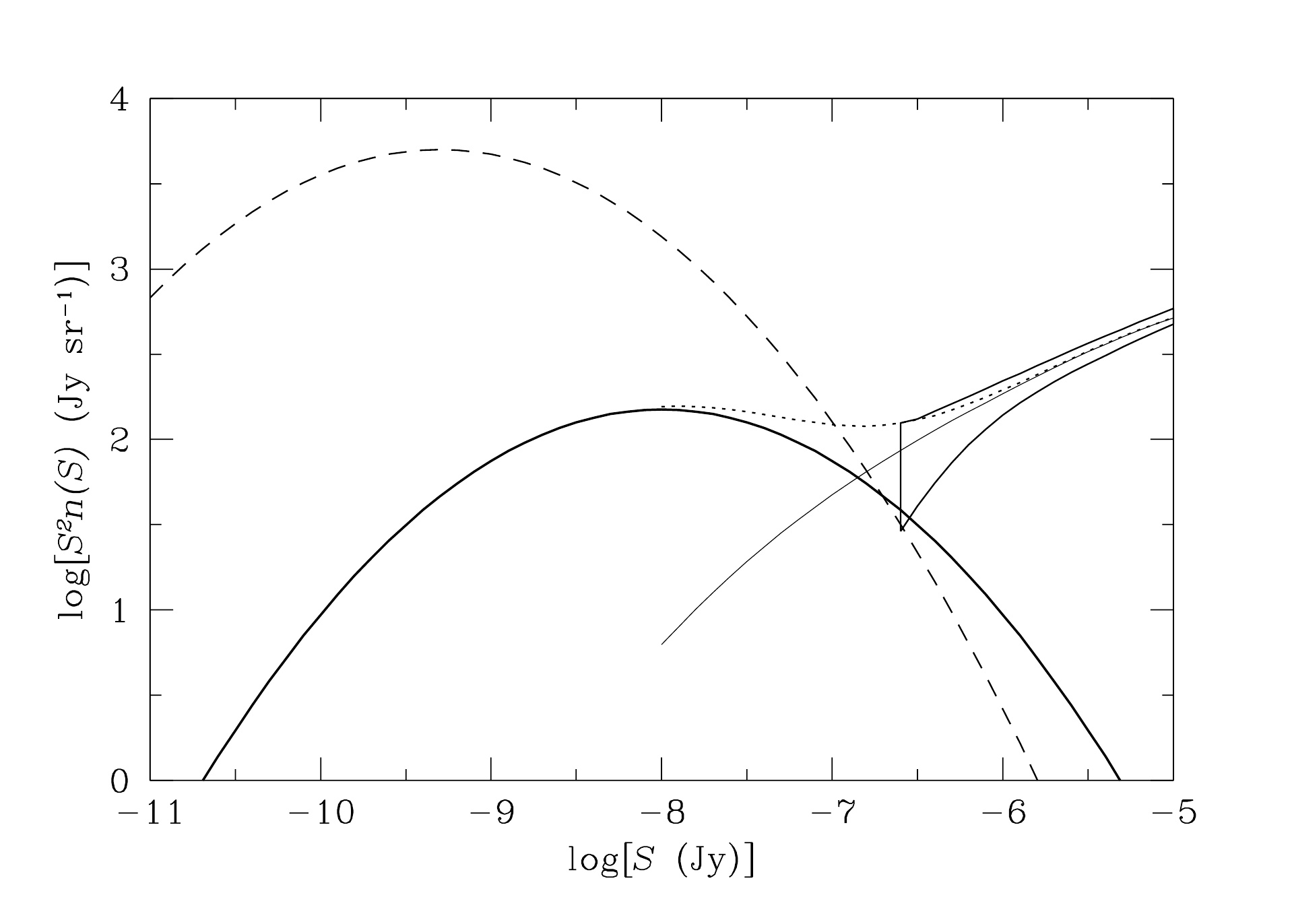}
  \caption{Source counts at 1.4\,GHz consistent with the DEEP2 $P(D)$
    distribution are shown by the thin black line surrounded by its
    $\pm 1\sigma$ error region.  A hypothetical new population with
    logarithmic FWHM $\phi=2$ and $A=150$ located at $\log(S_{\rm pk})
    = -8$ (thick black parabola) is consistent with DEEP2, but it
    contributes only 10\,mK to the radio source background.  A
    hypothetical population contributing the necessary 0.4\,K to agree
    with the background measured by ARCADE 2 must have $A \approx
    5000$ and $S^2 n(S)$ peaking at $S_{\rm pk} \leq 0.5$\,nJy (black
    dashed parabola) to remain consistent with our DEEP2 $P(D)$
    observation.  \label{fig:srccounts}}
\end{figure}

The MeerKAT correlation interferometer used to make the DEEP2 image
does not respond to backgrounds smooth on angular scales $\gg
\theta_{1/2} = 7\,\farcs 6$, and the resolution of ARCADE 2 is too
coarse to detect individual $<0.1\,\mu\mathrm{Jy}$  sources, so there is actually no
\emph{observational} tension between our results in
Section~\ref{sec:knownsources} and the ARCADE 2 background.  However,
the DEEP2 $P(D)$ distribution can set a lower limit to the \emph{number} of
faint sources not much larger than $\theta_{1/2} = 7\,\farcs 6 \approx
50\,\mathrm{kpc}$ in the redshift range $0.5 < z < 5$ that can produce
a $\Delta T_\mathrm{b} \sim 0.4\,\mathrm{K}$ background smooth enough
to be consistent with the DEEP2 $P(D)$ distribution.  Very numerous
faint sources contribute a nearly Gaussian $P(D)$ distribution similar
to the instrumental noise distribution.  A source population with rms
confusion not much larger than $\sigma_\mathrm{c} \approx (0.58^2 -
0.57^2)^{1/2}\,\mu\mathrm{Jy\,beam}^{-1} \sim 0.1\,\mu{\rm Jy
  \,beam^{-1}}$ is consistent with the uncertainty in the measured
DEEP2 rms noise.  Figure \ref{fig:srccounts} plots the
brightness-weighted source counts $S^2n(S)$ as a function of
$\log(S)$. The two broad peaks corresponding to star-forming galaxies
and AGN are well represented by the approximation \citep{condon12}
$\log[S^2n(S)] \approx a - b[\log(S) - \log(S_{\rm pk})]^2$ or
\begin{equation}\label{eq:taylorS2n}
  S^2n(S) \approx A\exp\left\{-4\ln(2)\frac{[\log(S) - \log(S_{\rm pk})]^2}
  {\phi^2}\right\},
\end{equation}
where $\phi$ is the logarithmic FWHM and $S_{\rm pk}$ is the flux
density of the $S^2 n(S)$ peak; $\log(S_{\rm pk}) \sim -5$ and
$\log(S_{\rm pk}) \sim -1$ for SFGs and AGNs,
respectively, while both populations are well described by $\phi = 2$.
Inserting Equation \ref{eq:taylorS2n} into Equation \ref{eq:tb} and
integrating over flux density determines the peak amplitude $A$
for FWHM $\phi$ for a new population adding $\Delta T_{\rm b}$ to the
total sky brightness:
\begin{equation}\label{eqn:amp}
  A\phi = \frac{4k_{\rm B}\nu^2}{\ln(10)c^2}
  \left[\frac{\ln(2)}{\pi}\right]^{1/2}\Delta T_{\rm b}.
\end{equation}
Equation~\ref{eqn:amp} is valid for any count peak flux density
$S_\mathrm{pk}$.  Because $\Delta T_{\rm b}$ fixes the product
$A\phi$, a new population with $T_{\rm b}$ either has a small number
of sources $N$ with a narrow FWHM $\phi$ but large amplitude $A$, or a
broad peak $\phi$ with a larger number of sources per square arcmin
and a fainter peak flux density.

The known AGN and SFG populations are well characterized by Gaussians
of FWHM $\phi = 2$, so we assumed $\phi = 2$ for the hypothetical new
population.  In order to explain the excess $\Delta T_\mathrm{b} = 0.4
\,\mathrm{K}$, the amplitude of this population must be $A \sim
5000\,{\rm Jy\,sr^{-1}}$.

For any $S_\mathrm{pk}$ we can find the maximum value of $A$ that is
consistent with the DEEP2 $P(D)$.  Inserting a new population of
sources with $\phi = 2$ and $\log(S_{\rm pk}) = -8$, we ran DEEP2
image simulations starting at $\log[S\mathrm{(Jy)}] = -10$ for
increasing values of $A$ in steps of 10 from $A = 0$.  They were
repeated for $\langle\sigma_\mathrm{n}\rangle = 0.56\,\mu{\rm
  Jy\,beam^{-1}}$ and $0.58\,\mu{\rm Jy\,beam^{-1}}$ in addition to
the measured rms noise averaged throughout the $P(D)$ region of radius
$500''$, $\langle\sigma_\mathrm{n}\rangle = 0.57\mu{\rm
  Jy\,beam^{-1}}$.  Having run a minimum of 50 iterations, we
determined the $\chi^2$ value of each simulated $P(D)$ distribution
(containing the new population of amplitude $A$) to the observed
$P(D)$.  The value of $A$ such that 16\% of the simulations had
$\chi^2 < \chi^2_{0}$ is the $1\sigma$ upper limit to the amplitude
$A$ that is consistent with our observed $P(D)$.

For rms noise values $\sigma_n = 0.56$ and $0.57\,\mu{\rm Jy\,beam^{-1}}$
the maximum amplitude consistent with the observed DEEP2 $P(D)$
distribution is $A \approx 65$. Simulations assuming rms noise
$\sigma_\mathrm{n}=0.55\,\mu{\rm Jy\,beam}^{-1}$ allow $A < 150$.  The
final $1\sigma$ upper bound on the counts of nJy sources is the
combination of Equation \ref{eq:bounds} and the curve representing the
sum of the new population and the measured counts from DEEP2 (dotted
curve in Figure \ref{fig:srccounts}).

Figure \ref{fig:srccounts} shows the brightness-weighted source counts
for the hypothetical population with $A=150, \phi=2$, and $S_{\rm
  pk}=10\,$nJy as the thicker black curve.  That new population adds
only $\Delta T_\mathrm{b} \sim 0.01 \,\mathrm{K}$ to the total radio
source background at 1.4 GHz, yet it must contain at least 3,000
sources per arcmin$^2$, exceeding by a factor of 30 the sky density of
galaxies brighter than $m_\mathrm{AB} +29$, the magnitude of the Large
Magellanic Cloud at redshift $z = 2$) in the Hubble Ultra Deep
Field \citep{beckwith06}.

A hypothetical new population contributing the full 0.4\,K ARCADE 2 excess 
background is consistent with the narrow DEEP2 $P(D)$ distribution
only if $\log(A) \approx 3.7$, the sources are randomly distributed on
the sky, and $S_{\rm pk}< 0.5 \,\mathrm{nJy}$ at 1.4\,GHz.
as indicated by the dashed parabola in Figure \ref{fig:srccounts}).
This upper limit to $S_{\rm pk}$ is a factor of ten lower than the
\citet{condon12} limit and more strongly excludes any bright population
of numerous faint sources that cluster like galaxies or parts of
galaxies.

\section{Summary}\label{sec:summary}

In this work, we presented source counts in the eight
  decades of flux density from $S = 0.25\,\mu\mathrm{Jy}$ to
  $S = 25\,\mu\mathrm{Jy}$ using the MeerKAT DEEP2 field and archival
  NVSS data.
  \begin{itemize}
    \item Statistical source counts betwen $S = 0.25\,\mu\mathrm{Jy}$
  and $S = 10\,\mu\mathrm{Jy}$ (Section~\ref{sec:pofd}) were estimated
  from the confusion $P(D)$ distribution within $500\arcsec$ of the DEEP2
  pointing center. Simulations of the radio sky were developed and used
  to constrain the counts and their uncertainties.
  \item We constructed a uniformly sensitive catalog of
$\approx 17,000$ discrete sources stronger than $S =
    10\,\mu\mathrm{Jy}$ at 1.266\,GHz
in $\Omega_{1/2} = 1.04\,\mathrm{deg}^2$
(Section~\ref{sec:deep2cat}) and used it to count discrete sources in
the flux-density range $10\,\mu\mathrm{Jy} \leq S < 2.5
\mathrm{\,mJy}$ (Section~\ref{sec:deep2count}).
\item The NVSS catalog of
radio source components was used to determine 1.4\,GHz source counts
between $S = 2.5\,\mathrm{mJy}$ and $S = 25\,\mathrm{Jy}$ in
$\Omega \approx 7.016 \, \mathrm{sr}$.
(Section~\ref{sec:nvsscount}).
  \end{itemize}
We find good agreement with previously published 1.4\,GHz counts, but report
higher source counts (at the $2\sigma$ level) than previously published 3\,GHz
counts from the VLA-COSMOS catalog. The agreement among all $\mu\mathrm{Jy}$
source counts is much improved from past studies.

Sources stronger than
our lower limit of $S = 0.25\,\mu\mathrm{Jy}$ resolve $>99\%$ of the AGN
contribution and $>96\%$ of the SFG contribution to the sky brightness
temperature.  The maximum source count amplitude for a hypothetical ``new''
population explaining the ARCADE 2 excess radio background is
$\log(A) \approx 3.7$, and the peak of the distribution must be fainter than
$S_{\rm pk} \sim 0.5\,\mathrm{nJy}$ to remain consistent with the DEEP2 $P(D)$
distribution. In a second paper (Matthews et al.,
in prep) we will use the results from this paper to estimate the star
formation history of the Universe.

\acknowledgments
We thank the anonymous referee for a careful reading of the paper and
especially constructive comments.
The MeerKAT telescope is operated by the South African Radio Astronomy
Observatory, which is a facility of the National Research
Foundation, an agency of the Department of Science and Innovation.
The National Radio Astronomy Observatory is a facility of the National
Science Foundation operated by Associated Universities, Inc.  This
material is based upon work supported by the National Science
Foundation Graduate Research Fellowship under Grant No. DDGE-1315231.
Support for this work was provided by the NSF through the Grote Reber
Fellowship Program administered by Associated Universities,
Inc./National Radio Astronomy Observatory.
This research has made use of the NASA/IPAC Infrared Science Archive,
which is funded by the National Aeronautics and Space Administration and
operated by the California Institute of Technology.\\

\facilities{Gaia, IRSA, MeerKAT}

\bibliographystyle{aasjournal}
\bibliography{paperv12}

\end{document}